\documentclass[aps,prx,reprint,superscriptaddress,letterpaper,twocolumn,longbibliography,floatfix]{revtex4-2}
\usepackage[english]{babel}
\usepackage{ucs}
\usepackage[utf8]{inputenc}
\usepackage{amsmath}
\usepackage{amssymb}
\usepackage{amsthm}
\usepackage{amsfonts}
\usepackage{mathtools}
\usepackage{mathrsfs}
\usepackage{dsfont} 
\usepackage{graphicx}
\usepackage{bm}
\usepackage{bbm}
\usepackage{empheq}
\usepackage{cases}
\usepackage{euscript}

\usepackage[svgnames,dvipsnames,x11names]{xcolor}
\usepackage[colorlinks=true,linkcolor=SpringGreen4,citecolor=blue,urlcolor=Magenta3]{hyperref}
\newcommand{\bra}[1]{\langle #1|}
\newcommand{\ket}[1]{|#1\rangle}
\newcommand{\braket}[2]{\langle #1|#2\rangle}
\newcommand{\ketbra}[2]{\ket{#1}\!\bra{#2}}

\newcommand{\mm}[1]{\mathrm{#1}}
\newcommand{\abs}[1]{\left|#1\right|}

\newcommand{\di}[1]{\mathop{}\!\mathrm{d} #1}

\def \ud{\mathrm{d}}

\def \uf{\mathrm{f}}

\def \rd{\partial}
\def \pv{\mbox{\boldmath$p$}}

\def \hH{\hat{H}}

\def \hU{\hat{U}}

\def \hS{\hat{S}}

\def \hW{\hat{W}}

\DeclareFontFamily{OT1}{pzc}{}
\DeclareFontShape{OT1}{pzc}{m}{it}{<-> s * [1.10] pzcmi7t}{}
\DeclareMathAlphabet{\mathpzc}{OT1}{pzc}{m}{it}

\begin{document}

\title{Topological Operations Around Exceptional Points via Shortcuts to Adiabaticity}

\author{Vishnu Chavva}
\affiliation{Department of Physics and Applied Physics, University of Massachusetts Lowell, Lowell, MA 01854, USA}

\author{Hugo Ribeiro}
\affiliation{Department of Physics and Applied Physics, University of Massachusetts Lowell, Lowell, MA 01854, USA}

\begin{abstract}
The existence of singularities in the spectrum of non-Hermitian Hamiltonians leads to a non-trivial spectral topology which can be
exploited to generate topological operations. However, their implementation has remained elusive due to the
difficulty of generating a true adiabatic evolution. Here, we develop fast, robust control protocols that generate 
a desired topological operation. Our strategy relies on shortcuts to adiabaticity, but is not a trivial
extension. The presence of spectral singularities renders the strategy developed for Hermitian Hamiltonians impractical as it will
lead to faulty control protocols. Moreover, due to the dynamics sensitivity to parameter uncertainties, not all shortcuts to
adiabaticity can be used in a realistic setting. We illustrate our method in the context of a two-mode non-Hermitian Hamiltonian
and discuss why in general celebrated shortcuts to adiabaticiy like transitionless driving and superadiabatic transitionless are
not appropriate control protocols for non-Hermitian systems. 
\end{abstract}

\maketitle

\section{Introduction} 
\label{sec:intro}

\begin{figure}[t]
    \centering
    \includegraphics[width=\columnwidth]{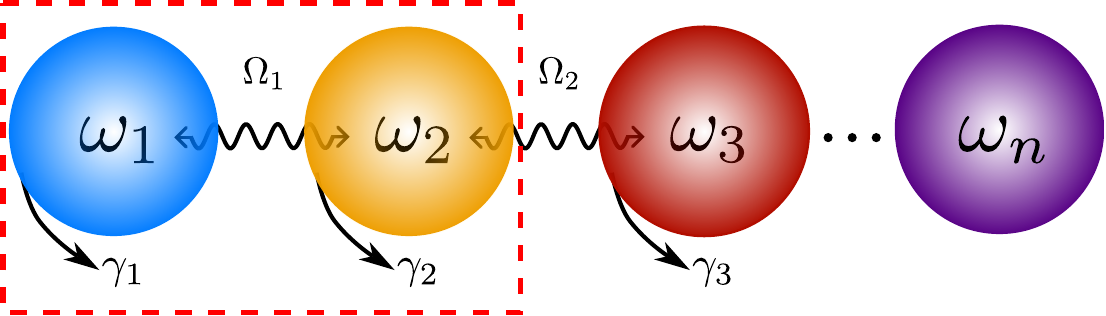}
    \caption{N-coupled modes with local dissipation. The dissipative evolution of the system can be described by an effective
    non-Hermitian Hamiltonian with corresponding coupling strengths $\Omega_\mm{n}$ and dissipation rates $\gamma_\mm{n}$.}
    \label{fig:intro_fig} 
\end{figure}

One of the fundamental postulates of quantum mechanics is that closed systems are described by Hermitian Hamiltonians
($\hH=\hH^\dagger$). This guarantees a real spectrum for $\hH$ and ensures that the evolution operators obtained by solving
Schrödinger's equation are unitary and probability is conserved. However, most physical systems are best modeled by assuming that
they exchange energy with a large environment.  A straightforward way of modeling such dissipative systems is through the use of
non-Hermitian (NH) Hamiltonians ($\hH\neq \hH^\dagger$) (see Fig.~\ref{fig:intro_fig}). 

Theories based on NH Hamiltonians have been successful in describing photonic systems exhibiting
gain/loss~\cite{feng2017,Li2023,Mohammad2019}, interactions between atomic systems and their
environment~\cite{Zhang2018,dum1992,wang2022}, and non-equilibrium dynamics present in condensed matter
systems~\cite{McDonald2022,Li2024,Xiao2024}.  The unique properties of these NH Hamiltonians have, and continue to be, the driving
force behind their growing theoretical and experimental interest~\cite{ding2022,ashida2020,decarlo2022,Wang2023}.

A signature feature of NH Hamiltonians is their complex spectrum. Generally, the real part of the spectrum corresponds to the
system energy while the imaginary part accounts for the dissipative behavior. If the NH Hamiltonian depends on externally
controllable parameters, one can tune the interplay between dissipative and coherent dynamics. Furthermore, for certain parameter
values, eigenvalues and eigenvectors coalesce~\cite{Kato1976,ashida2020,heiss2012,Mohammad2019,Ozdemir2019,okuma2023}, and the NH
Hamiltonian cannot be diagonalized.  These singularities in parameter space are known as exceptional points (EPs) and give rise to
a non-trivial spectral topology~\cite{dembowski2001,Guria2024,Ding2016}.  

The archetypal characteristic is a spectrum that does not return to itself when the paramaters are varied along a closed contour
(in parameter space) that encircles an EP. Eigenvalues appear to ``swap'' with one another, forming a so-called eigenvalue braid.
This topological behavior is appropriately referred to as eigenvalue braiding~\cite{patil2022,Uzdin2011,Wojcik2022}, and
constitutes much of the interest in this work. Additionally, as the eigenvalues braid, the eigenvectors acquire an invariant
topological phase, determined by the dimensions of the NH Hamiltonian~\cite{berry1984,gong2018,mailybaev2005_2,liang2013}. 

Realizing eigenvalue braiding, however, is a cumbersome process since only one eigenstate (the least damped eigenstate) follows
adiabatic evolution ~\cite{liu2021,doppler2016,Chen2021,Wang2018,Nenciu1992,Moiseyev2011}. Thus, initializing an instantaneous
eigenstates of a non-Hermitian and ``slowly'' varying its parameters does not generate an evolution that guarantees eigenvalue
braiding~\cite{kato1950,Berry2011,Mustapha2017}.

In this work, we present a highly general strategy to design fast, robust control protocols that generate eigenvalue braiding for
NH systems. Our approach takes inspiration from the field of shortcuts-to-adiabaticity
(STAs)~\cite{demirplak2003,demirplak2008,berry2009,Odelin2019,Hatomura2024,chen2012}, specifically the dressed state
method~\cite{baksic2016,Zhou2017}. While it would appear that extending the dressed state method to NH Hamiltonians would be a
straightforward generalization, the non-trivial spectral topology of NH Hamiltonians, and in particular the presence of
singularities, prohibits one from following the framework developed for Hermitian Hamiltonians. Doing so will lead to control protocols
that fail to mimic the desired adiabatic evolution. 

Moreover, and in stark contrast to Hermitian STAs, non-Hermitian STAs do not inherit the robustness of the original adiabatic
protocol against parameter uncertainties. This follows from the violation of the adiabatic theorem for NH Hamiltonians.  Our work
shows how to construct adequate dressed states that allow one to find robust, experimentally realizable control schemes.  Our
method is independent of a particular choice of contour in parameter space and, much like the Hermitian dressed state method,
offers considerable freedom to design protocols that obey the constraints of specific platforms.

Extending STAs to NH Hamiltonians is not a new idea. Several works~\cite{ibanez2011,torosov2013} have already discussed how to
realize transitionless driving (TD)~\cite{demirplak2003,berry2009} in the context of NH Hamiltonians. However, these work do not
discuss how to restore adiabaticity around EPs nor the robustness of the non-Hermitian TD against parameter uncertainties.
Parameter uncertainties effectively lead to transitions between (unperturbed) eigenstates, including transitions to the least
damped eigenstate. Populating the latter will unavoidably result in its preparation at the end of the protocol, independently of
which eigenstate was initially prepared~\cite{ribeiro2021}. More recently, two works~\cite{Arkhipov2024,wu2024} showed that there
exists a special choice of closed contour around an EP that allows one to generate adiabatic evolution akin to that of Hermitian
systems. While the strategy allows one to realize and study adiabatic passage with NH Hamiltonians, it remains constrained to one
specific class of trajectories. As we discuss, our method is not restricted by a choice of initial contour.  This allows one to
explore adiabatic phenomena around EPs without being limited to a single type of evolution.

The reminder of this paper is organized as follows:~In Sec.~\ref{sec:prelim} we discuss how singularities in the spectrum of NH
Hamiltonians must be handled to define valid STAs and how the resulting framework fundamentally differs from Hermitian STAs. In
Sec.~\ref{sec:two-mode}, we apply our formalism to a two-mode system to illustrate the differences between the Hermitian and
non-Hermitian case, to discuss the issue of robustness, and to advance NH STAs that are both robust against parameter
uncertainties and experimentally feasible.

\section{Preliminaries}
\label{sec:prelim}

\subsection{Spectral flow and topological operations}
\label{subsec:TopOp}

We consider the generic $N$-dimensional non-Hermitian Hamiltonian 
\begin{equation}
    \hH= \hH(\pv),
    \label{eq:GenNonHermitianH}
\end{equation}
where $\pv = (p_1,\dots,p_n) \in \mathbb{R}^n$ is a shorthand notation to indicate that $\hH$ depends on $n$ parameters that can be externally
varied. The right eigenstates of Eq.~\eqref{eq:GenNonHermitianH} and corresponding right eigenvalues are defined by 
\begin{equation}
    \hH(\pv) \ket{\psi_j (\pv)} = \lambda_j (\pv) \ket{\psi_j (\pv)}.
    \label{eq:InstEigStates}
\end{equation}
Generically, as one varies $\pv$, the spectrum $\mm{Sp}[\hH(\pv)] = \{\lambda (\pv) | \mm{det}[\hH(\pv) - \lambda(\pv)
\mathbbm{1}] =0\} \equiv \{\lambda_j (\pv)\}_{j=1}^N$ becomes degenerate at singularities known as exceptional points of order $k$
($\mm{EP}_k$), with $2 \leq k \leq N$~\cite{Kato1976,gilmore2007}. At these degeneracy points $k$ eigenvalues and eigenstates
coalesce and $\hH$ becomes non-diagonalizable. Since the eigenvalues of a non-Hermitian Hamiltonian are in general complex (the
spectrum of $\hH$ can be entirely real, if parity-time symmetry is respected\cite{Bender1998,Ding2015,Klauck2019}), an $\mm{EP}_k$
formally corresponds to a branch point (endpoint of a branch cut) associated to the function $\lambda_j\,:\,\mathbb{R}^n \to
\mathbb{C}$, $\pv \mapsto \lambda_j(\pv)$~\cite{ding2022}.

We label the position in parameter space of the $l$th $\mm{EP}_k$ by $\pv_{k,l}$ and define $\mathit{B}_k=\{\pv_{k,l}\}_{l=1}^L$
as the parameter subspace formed by all $\mm{EP}_k$'s. The union of all $\mathit{B}_k$'s forms the parameter space
$\mathit{B}_\mm{deg} = \cup_k \mathit{B}_k$ where the spectrum of Eq.~\eqref{eq:GenNonHermitianH} is degenerate. We denote the
complement of $\mathit{B}_\mm{deg}$ as $\mathit{B}_\mm{ng} = \mathit{B}_\mm{deg}^\complement$, which corresponds to the parameter
space where the spectrum is non-degenerate. The images of $\mathit{B}_k$ $\forall k$ and $\mathit{B}_\mm{ng}$ under the action of
$\mm{Sp}[\hH(\pv)]$ define topological manifolds that are in general
non-trivial~\cite{Guria2024,patil2022,Lee2012,Demange2012,Graefe2008,Ding2015,Ryu2012,Ding2016}, and which we denote as
$\mm{Sp}(\mathit{B}_k)$ and $\mm{Sp}(\mathit{B}_\mm{ng})$, respectively.

The non-trivial topology of $\mm{Sp}(\mathit{B}_\mm{ng})$ becomes apparent when one studies how the spectrum varies along a family
of closed contours (loops) with a same basepoint. We denote such a family of loops by $\gamma_s [\pv(\varepsilon)] \subset
\mathit{B}_\mm{ng}$ with $\varepsilon \in [0,1]$. Since any loop $\gamma_s$ starts and ends at the same basepoint, we have
$\hH[\gamma_s (\varepsilon =0)] = \hH[\gamma_s (\varepsilon =1)]$ and therefore the spectrum must return to itself at the end of
the loop.  However, it might do so in a non-trivial fashion according to
\begin{equation}
    \lambda_j [\pv (\varepsilon =0)] \to \lambda_j [\pv (\varepsilon =1)] = \lambda_{\sigma (j)} [\pv (\varepsilon =0)],
    \label{eq:EigValPermutation}
\end{equation}
where $\sigma_n(j)$ is one of $1\leq n \leq N!$ permutations acting on the set $\{1,\ldots,j,\ldots,N\}$, and which is solely
determined by the choice of loop $\gamma_s$. 

Pairs of loops that are homotopic (can be continuously deformed into one another) in $\mathit{B}_\mm{ng}$ will lead to the same
permutation $\sigma_n$. By defining $\pi_1 (\mathit{B}_\mm{ng})$ as the set of homotopy equivalence classes of loops and
considering the concatenation of loops as a binary operation on $\pi_1 (\mathit{B}_\mm{ng})$, we can make $\pi_1
(\mathit{B}_\mm{ng})$ into a group called the fundamental group\cite{Hatcher2002}. Each element of $\pi_1$ corresponds to a unique
permutation $\sigma_n$ and the group operation between two elements of $\pi_1$ is associated to the composition of permutations.

To understand how to characterize the elements of $\pi_1 (\mathit{B}_\mm{ng})$, it is useful to consider the image of a loop under
the action of $\mm{Sp}[\gamma(\pv)]=\{\lambda_j [\pv (\varepsilon)]\}_{j=1}^N$. This produces a contour on
$\mm{Sp}(\mathit{B}_\mm{ng})$ that formally corresponds to a braid of $N$ strands [see Fig.~\ref{fig:STA_map}(a)], commonly known
as an eigenvalue braid~\cite{dembowski2001,Berry2011,Uzdin2011,heiss2000,Berry1998}. Since $\mm{Sp}[\gamma(\pv)]$ defines an
isotopy equivalence class of braids (pairs of braids are isotopic, if they can be continuously deformed into one another while
keeping the end points fixed and the strands from intersecting) and one can concatenate braids, we can define a group whose
elements are equivalence classes of braids with $N$ strands. This group is known as the Artin group $B_N$\cite{Artin1947}. An
element of $B_N$ corresponds to an element of $\pi_1 (\mathit{B}_\mm{ng})$, since two isotopic braids arise from two homotopic
loops $\gamma$. This implies $\pi_1 (\mathit{B}_\mm{ng}) \cong B_N$.

Since braid concatenation is in general not a commutative operation, one can generate braids belonging to different
isotopy classes by concatenating two same loops in a different order. This allows one to represent eigenvalue braiding as a
permutation operation defined by
\begin{equation}
    \hat{P}_{\sigma_n}= \sum_{j=1}^N \exp[i \beta_{\sigma_n(j),j}] \ketbra{\psi_{\sigma_n (j)} [\pv(\varepsilon=0)]}{\psi_j[\pv(\varepsilon=0)]},
    \label{eq:EigVecPermutation}
\end{equation}
and which in general fulfills the condition $[\hat{P}_{\sigma_n}, \hat{P}_{\sigma_m}]\neq 0$, if $\sigma_n \neq
\sigma_m$. Generating operations that correspond to $\hat{P}_\sigma$ to control a system would be beneficial, as these
operations are robust against imprecisions always present in control schemes, e.g., amplitude and phase of driving fields.

\subsection{``Adiabatic'' Evolution and Holomorphic Change of Frame Operators}
\label{subsec:prelim_dynamics}

Generating a topological operation corresponding to a particular $\hat{P}_{\sigma_j}$ requires adiabatically evolving the system by
varying $\hH$ along a control loop $\gamma$ that varies in time. Assuming the equation of motion to be linear and recasting it in
the form of a Schrödinger equation, we have that the flow operator $\hat\Phi (t)$ [equivalent to the unitary evolution operator
$\hU (t)$] obeys
\begin{equation}
    i \rd_t \hat\Phi (t) = \hH(t) \hat\Phi (t),
    \label{eq:SchrodingerEq}
\end{equation}
with $\hat\Phi (0) = \mathbbm{1}$. The flow $\hat\Phi (t)$ allows one to find the state of the system at time $t$,
if one knows the state at time $t=0$ through the relation $\ket{k (t)} = \hat\Phi(t) \ket{k (0)}$. The problem now is to
engineer an $\varepsilon (t)$ such that $\hH\{\gamma [\varepsilon(t/t_\uf)]\}$ generates an adiabatic dynamics leading to $\hat\Phi
(t_\uf) = \hat{P}_\sigma$. This is, however, impossible since it has been shown that systems governed by a non-Hermitian
Hamiltonian violate the adiabatic theorem~\cite{xu2016,nenciu1980,Uzdin2011,Berry2011}.

This is best understood in the frame of instantaneous eigenstates (adiabatic frame), where $\hH(t)$ is diagonal at
each instant of time $t$. In the adiabatic frame, the non-Hermitian Hamiltonian governing the dynamics is given by
\begin{equation}
    \begin{aligned}
        \hH_\mm{ad} (t) &= \hS_\mm{ad}^{-1} (t) \hH(t) \hS_\mm{ad}(t) - i \hS_\mm{ad}^{-1} (t) \rd_t\hS_\mm{ad} (t)\\
        &= \sum_j \lambda_j (t) \ketbra{\psi_j}{\psi_j}- i \hS_\mm{ad}^{-1} (t) \sum_{j=1}^n \frac{\rd \hS_\mm{ad}(t) }{\rd \pv_j}
        \dot{\pv_j} (t),
    \end{aligned}
    \label{eq:GenericTimeDepH}
\end{equation}
where we use Newton's notation for time differentiation, $\dot{x} (t) = \mm{d}x (t)/\mm{d}t$ and 
\begin{equation}
    \hS_\mm{ad}[\pv(t)] = \sum_{j=1}^{n} \ketbra{ \psi_j [\pv(t)]}{\psi_j}.
    \label{eq:GenSAd}
\end{equation}
is the time-dependent change of frame operator leading to $\hS_\mm{ad}^{-1} (t) \hH(t) \hS_\mm{ad}(t)$ being
diagonal.

Since Eq.~\eqref{eq:GenSAd} is explicitly time-dependent (the instantaneous eigenstates of $\hH$ are not stationary), the
non-Hermitian Hamiltonian in the adiabatic frame picks up a non-inertial term, which couples the instantaneous eigenstates [see
Eq.~\eqref{eq:GenericTimeDepH}]. This coupling leads to transitions between the instantaneous eigenstates, which in contrast to
the Hermitian case, prevent the system from following adiabatic dynamics even in the long-time limit [see
Fig.~\ref{fig:STA_map}(b)], i.e., when $t_\uf \to \infty$. The violation of the adiabatic theorem for non-Hermitian systems has
been observed experimentally, particularly in two-mode
systems~\cite{Li2023,zhong2018,dai2024,elganainy2018,Yang2024,delPino2022,Lau2018,Huang2024}.

Additionally, to correctly define Eq.~\eqref{eq:GenericTimeDepH} one must construct a change of frame operator $\hS_\mm{ad}(t)$
that is holomorphic everywhere except at the location of the EPs. This is necessary since Eq.~\eqref{eq:GenericTimeDepH} requires
one to calculate the partial derivative of $\hS_\mm{ad} (\pv)$ with respect to the parameters that define the contour, which is
only possible if the matrix elements of $\hS_\mm{ad} (\pv)$ are holomorphic functions. 

It is always possible to construct a holomorphic change of frame operator by first defining the eigenvalues $\lambda_j (t)$ on
their associated Riemann surface~\cite{churchill2009} and then solving for the eigenstates $\ket{\psi_j[\pv(t)]}$, which define
Eq.~\eqref{eq:GenSAd}. We show an explicit example of this procedure in Sec.~\ref{sec:two-mode}.

\subsection{Dressed State Approach to Non-Hermitian Shortcuts-to-Adiabaticity}
\label{subsec:STAs}

\begin{figure}[t]
    \centering
    \includegraphics[width=\columnwidth]{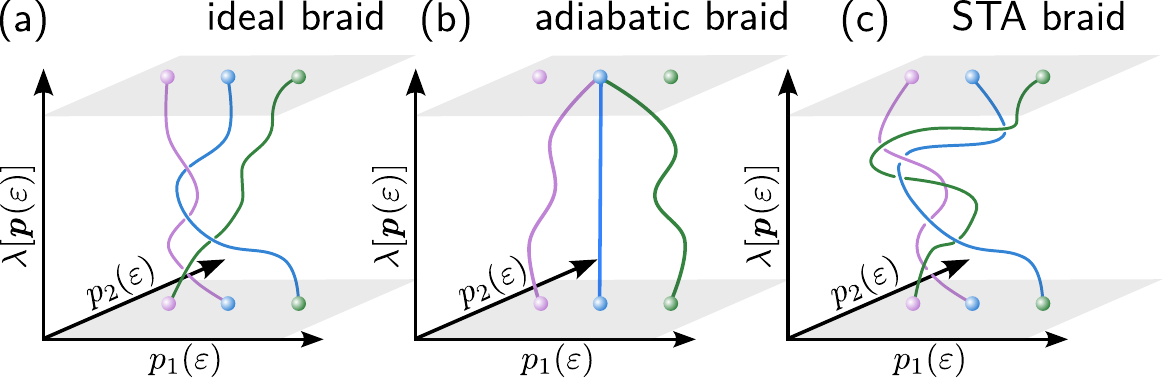}
    \caption{
        Schematic representations of the eigenvalues braids traced by a choice of contour $\gamma
        [\pv(\varepsilon)]$ with $\pv=(p_1,p_2)$. (a) Ideal braid obtained from spectral flow. (b) Adiabatic evolution of
        eigenstates does not yield the desired eigenvalue braiding. (c) STAs do not in general
        produce the ideal braid, but still lead to the desired topological operation. 
    } 
    \label{fig:STA_map} 
\end{figure}

Now that we have laid down the fundamental concepts allowing one to find the correct representation of any time-dependent
non-Hermitian operator in the adiabatic frame, we can formulate a theory to construct non-Hermitian STAs. 

Our approach builds on the ideology of the dressed state approach for Hermitian systems~\cite{baksic2016}:~If one is only
interested in the topological operation generated by encircling an EP with a control loop $\gamma[\pv(t)]$ of duration $t=t_0$,
then it is not necessary to follow the ideal adiabatic evolution at all times. It is sufficient to generate an evolution that only
resembles the adiabatic evolution at the initial $t=0$ and final time $t=t_0$. In other words, we ensure only that the end points
of the eigenvalue braids generated by the STA correspond to the end points of the ideal eigenvalue braids [see
Fig.~\ref{fig:STA_map}(c)]. 

Like for Hermitian systems, we must (1) choose a dressing transformation $\hS_\mm{dr}(t)$ for the instantaneous eigenstates which
vanishes at $t=0$ and $t=t_0$, and (2) find the associated non-Hermitian control Hamiltonian $\hW (t)$ such that the modified
non-Hermitian Hamiltonian $\hH_\mm{mod} (t) = \hH(t) + \hW (t)$ generates the desired evolution at $t=t_0$.  In addition to these
requirements, we must introduce a new constraint:~(3) the change of frame operators must be holomorphic functions along
$\gamma(t)$, otherwise the non-Hermitian STAs will be unusable.  

A generic dressing transformation can be parametrized by
\begin{equation}
    \hS_\mm{dr} [\pv(t)] = \sum_{j=1}^{n} \ketbra{ \phi_j [\pv(t)]}{\phi_j},
    \label{eq:GenSdrSTA}
\end{equation}
where $\ket{\phi_j}$ are the dressed eigenstates. With this parametrization condition (1) implies that Eq.~\eqref{eq:GenSdrSTA}
must satisfy 
\begin{equation} 
    \hS_\mm{dr} (0) = \hS_\mm{dr}(t_0) = \mathbbm{1}.
    \label{eq:STA_condition}
\end{equation}
This guarantees that the dressed states coincide with the instantaneous eigenstates $\ket{\psi_j}$ at $t=0$ and $t=t_0$. This is 
equivalent to saying that $\hW(t) = 0$ at the end points, which implies that the spectrum of $\hH_\mm{mod} (t)$ reduces to the
spectrum of $\hH(t)$ at $t=0$ and $t=t_0$. 

In the dressed frame, the dynamics is described by the non-Hermitian Hamiltonian
\begin{equation} 
    \hH_\mm{dr} (t) = \hS_\mm{dr}^{-1} (t) \hH_\mm{ad} (t) \hS_\mm{dr}(t) - i \hS_\mm{dr}^{-1} (t) \rd_t\hS_\mm{dr} (t).
    \label{eq:H_dr}
\end{equation}
By imposing
\begin{equation} 
    \bra{\phi_j} \hH_\mm{dr} (t) \ket{\phi_i} = 0
    \label{eq:H_dr_condition}
\end{equation}
for all $i \neq j$ allows one to find the matrix element of the control operator $\hW (t)$. 

Since the dressing transformation is explicitly time-dependent, $\hH_\mm{dr} (t)$ also contains a non-inertial term [see
Eq.~\eqref{eq:H_dr}] obtained from the derivative of $\hS_\mm{dr} (t)$. To obtain the correct non-inertial term, which defines
$\hW (t)$ [see Eq.~\eqref{eq:H_dr_condition}], one must construct a dressing transformation which is holomorphic along $\gamma
(t)$.  Therefore, one can see the necessity of including requirement (3) in defining the criteria for a non-Hermitian STA.

As we discuss in Sec.~\ref{sec:SATD}, this leads to substantial differences between the Hermitian and non-Hermitian case. In the
Hermitian case, any choice of dressing can be viewed as a family of transformations labeled by the loop time $t_0$ and will yield
a family of control Hamiltonians $\hW (t)$ also labeled by $t_0$. This does not hold true in the non-Hermitian case: A dressing
transformation that yields a physical $\hW (t)$ for one value of $t_0$ might yield an unphysical $\hW (t)$ for another value of
$t_0$. 

\section{Example:~Two-mode system} 
\label{sec:two-mode}

To make things concrete, and without loss of generality, we consider a symmetrized
two-mode non-Hermitian Hamiltonian
\begin{equation}
    \hH_\mm{sym} = -\left(\Delta + i\frac{\Gamma}{2}\right) \hat{\sigma}_{z} + \Omega \hat{\sigma}_{x},
    \label{eq:H_sym_static}
\end{equation}
that describes two coupled harmonic modes subject to damping [see Fig.~\ref{fig:intro_fig}~(a)]. Here, $\Delta= (\omega_1 -
\omega_2)/2$, is the frequency detuning, $\Omega$ is the coupling strength, and $\Gamma = (\gamma_1-\gamma_2)/2 > 0$ is the
difference between the damping rates associated to the individual modes. 

Equation~\eqref{eq:H_sym_static} depends on three parameters, which using the language of the previous section, we label by the
vector $\pv=(\Delta,\Omega,\Gamma) \in \mathbb{R}^3$. In spite of its simplicity, Eq.~\eqref{eq:H_sym_static} describes several
state-of-the-art experimental platforms~\cite{doppler2016,wilkey2023,chen2017,Bliokh2019,Yang2024,Lau2018,xu2016} where
non-Hermitian STAs can be readily implemented.

\subsection{Holomorphic change of frame operator}
\label{sec:2ModeHolomorphicS}

We start by showing how to find a holomorphic change of frame operator that diagonalizes Eq.~\eqref{eq:H_sym_static}. The
spectrum of $\hH_\mm{sym}$ is given by $\mm{Sp}[\hH_\mm{sym}(\pv)] = \{\pm \sqrt{(\Delta + i \Gamma/2)^2 + \Omega^2}\}$ and has
$\mm{EP}_2$'s located in parameter space determined by the system of equations
\begin{equation}
    \begin{aligned}
        -\frac{\Gamma^2}{4} + \Omega^2 + \Delta^2 &= 0, \\
        \Delta \Gamma &= 0.
    \end{aligned}
    \label{eq:EP2}
\end{equation}
The set of vectors $\pv_{2,l}= (\Delta_l,\Omega_l,\Gamma_l)$, $l\geq 1$, obeying Eq.~\eqref{eq:EP2} form the
parameter subspace $B_2$ where the spectrum is doubly degenerate. Since the spectrum of Eq.~\eqref{eq:H_sym_static} only has
$\mm{EP}_2$'s, we have $B_\mm{deg} = B_2$ and the subspace of parameters where the spectrum is non-degenerate is given by
$B_\mm{ng} = \mathbb{R}^3 \setminus B_2$.

\begin{figure*}[t] 
    \centering
    \includegraphics[width=2\columnwidth]{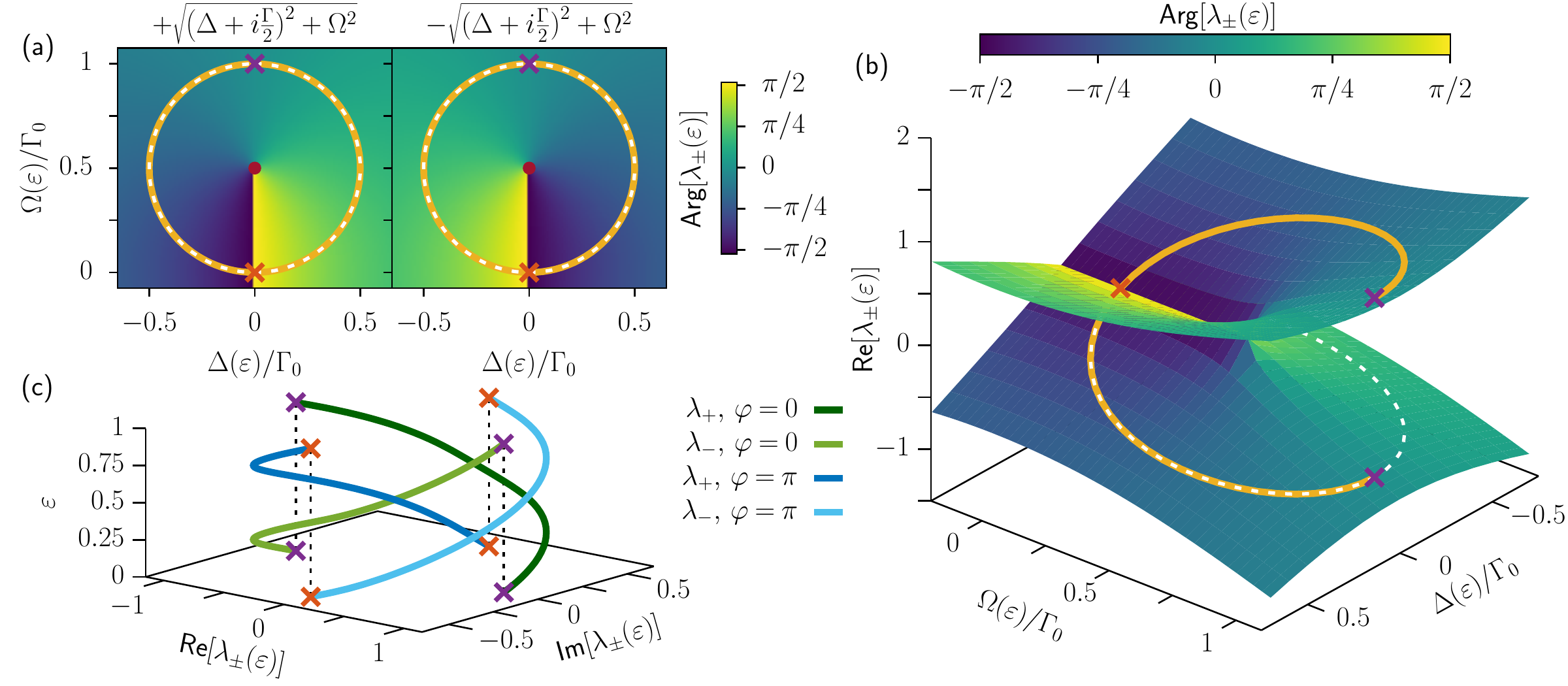}
    \caption{Riemann surface and topology of the square root function. (a) The argument of the eigenvalues of
        Eq.~\eqref{eq:H_sym_static} in the vicinity of the EP (red dot). The yellow (white dashed) circle represents a
        circular contour $\gamma_\mm{circ} (\varepsilon)$ with basepoint $\varphi=0$ indicated by the purple cross ($\varphi=\pi$, orange cross) [see
        Eq.~\eqref{eq:Num_delta_Num_omega}]. The argument is discontinuous along the contour. (b) Defining the eigenvalues of the NH
        Hamiltonian on their associated Riemann manifold leads to holomorphic functions along  $\gamma_\mm{circ} (\varepsilon)$.
        Contours with different basepoints lead to different spectral flows [yellow and white dashed traces correspond to contours
        defined in (a)]. (c) Eigenvalues braiding associated to contours defined in (a).
    }
    \label{fig:manifold}
\end{figure*}

Since the eigenvalues of $\hH_\mm{sym}$ behave like the complex square root function, they are not holomorphic on all
of $B_\mm{ng}$. The region defined by
\begin{equation}
    \begin{aligned}
        -\frac{\Gamma^2}{4} + \Omega^2 + \Delta^2 &\leq 0, \\
        \Delta \Gamma &= 0,
    \end{aligned}
    \label{eq:cross_condition}
\end{equation}
must be excluded as any $\pv$ belonging to this region is mapped to the interval $(-\infty,0]$ which corresponds to the branch cut
of the square root function [see Fig.~\ref{fig:manifold}(a)]. This shows the necessity of finding a different representation for
the eigenvalues since any closed contour $\gamma$  winding around $B_2$ necessarily intersects the branch cut region leading to a
phase shift of the eigenvalues [see Fig.~\ref{fig:manifold}~(a)]. This prevents one from finding a change of frame operator that
is holomorphic in $B_\mm{ng}$.

To overcome this issue, we express the eigenvalues of Eq.~\eqref{eq:H_sym_static} on the Riemann surface for the square
root function \cite{AhlforsSario1960} [see Fig.~\ref{fig:manifold}~(b)]. We have
\begin{equation}
    \lambda_\pm (\pv)= \pm \lambda (\pv) = \pm\cos\left[\chi (\pv)\right]\sqrt{\left(\Delta + i \frac{\Gamma}{2} \right)^{2} + \Omega^{2}},
    \label{eq:cont_lambda_static}
\end{equation}
where $\chi$ is a function that geometrically accounts for the ``gluing'' of the two leaves of the square root function that make
up the Riemann surface. Basically, the function $\chi$ is constructed to compensate for the $\pi$-phase shift incurred when
crossing the branch cut such that there is a seamless transition from one leaf of the square root function to the other.

Using Eq.~\eqref{eq:cont_lambda_static} we can find the right eigenmodes of $\hH_\mm{sym}$ [see Eq.~\eqref{eq:InstEigStates}]
such that the coefficients of their decomposition are holomorphic functions as
desired. We find 
\begin{equation}
    \ket{\psi_j (\pv)} = \left[-\left(\Delta + i\frac{\Gamma}{2}\right) + \lambda_j (\pv) \right] \ket{0} + \Omega \ket{1},
    \label{eq:StaticEigVecs}
\end{equation}
with  $j\in\{+,-\}$. This immediately yields the change of frame operator that diagonalizes Eq.~\eqref{eq:H_sym_static}, and which
is holomorphic on $B_\mm{ng}$,
\begin{equation}
    \hS_\mm{ad}(\pv) = \sum_{j=-,+} \ketbra{\psi_j}{\psi_j (\pv)},
    \label{eq:ChangeOfFrameStatic}
\end{equation}
where $\bra{\psi_j (\pv)}$ is a left eigenvector of $\hH_\mm{sym}$. Left eigenvectors obey the eigenvalue equation $\bra{\psi_j}
\hH_\mm{sym} = \lambda_j \bra{\psi_j}$ and are related to the right eigenvectors by the orthogonality relation
$\braket{\psi_k}{\psi_j} = \delta_{k,j}$, where $\delta_{k,j}$ is the Kronecker delta function. In the frame that diagonalizes
Eq.~\eqref{eq:H_sym_static} the eigenvectors $\ket{\psi_j}$ are independent of $\pv$ (constant eigenvectors).

Finally, we note that for non-Hermitian Hamiltonians with special symmetry like Eq.~\eqref{eq:H_sym_static}, it is often possible
to find the change of frame operator without having to explicitly find the eigenvalues and eigenvectors of $\hH$. Here, by making
the analogy to a spin-$1/2$ in an external magnetic field, the change of frame operator that diagonalizes the Hamiltonian can be
found by aligning the quantization axis with the orientation of the spin (rotation of the coordinate system).  We find that the
holomorphic change of frame operator can also be written as
\begin{equation}
    \hS_\mm{ad}(\pv) = \exp \left[-\frac{i}{2} \theta (\pv) \hat{\sigma}_y\right],
    \label{eq:S_ad_holo}
\end{equation}
with
\begin{equation}
    \theta (\pv) = -2\arctan\left(\frac{\Omega}{\Delta + i \frac{\Gamma}{2}+\sqrt{\left(\Delta + i \frac{\Gamma}{2} \right)^{2} +
    \Omega^{2}}}\right) + \chi (\pv).
    \label{eq:cont_theta_static}
\end{equation}

We have defined the pseudo-rotation angle $\theta (\pv)$ using the half-angle identity $\mm{arctan}[y/x] =
2\mm{arctan}[y/(x+\sqrt{x^2+y^2})]$ to ensure $\chi$ is the same function as the one introduced in
Eq.~\eqref{eq:cont_lambda_static}. In this case, $\chi$ ``glues'' together the leaves of the complex $\mm{arctan}$ function, which
has a branch cut on the interval $(-i\infty,-i] \cup [i,i\infty)$, to ensure that $\theta (\pv)$ is holomorphic on $\mathbb{R}^3
\setminus B_\mm{ng}$. 

Performing the similarity transformation defined by Eq.~\eqref{eq:S_ad_holo} leads to
\begin{equation}
    \hH_{\mm{diag}} (\pv) =  \hS_\mm{ad}^{-1} (\pv) \hH_{\mm{sym}} (\pv)  \hS_\mm{ad}(\pv) =
    \lambda (\pv) \hat{\sigma}_{z,\mm{diag}}.
    \label{eq:H_ad_static}
\end{equation}
with $\lambda (\pv)$ given by Eq.~\eqref{eq:cont_lambda_static}, which immediately yields the holomorphic eigenvalues.

\subsection{Topological operations with a two-mode system}

We are now in a position to understand what kind of topological operations one can generate by winding around $B_2$. To this end,
we consider a closed, circular contour $\gamma_\mm{circ}(\varepsilon)$ centered at $\pv_\mm{center} = (0, \Omega_0, \Gamma_0)$
with radius $\Delta_0$ parameterized by
\begin{equation}
    \begin{aligned}
        \Delta(\varepsilon)&= \Delta_0 \sin (\varepsilon + \varphi),\\
        \Omega(\varepsilon)&= \Omega_0+\Delta_0 \cos (\varepsilon + \varphi),\\
        \Gamma (\varepsilon) &= \Gamma_0,
    \end{aligned}
    \label{eq:Num_delta_Num_omega}
\end{equation}
with $\varepsilon \in [0,1]$. The basepoint of $\gamma_\mm{circ}(\varepsilon)$ can be unambiguously located in parameter space by
specifying the center of the circle [see red dot in Fig.~\ref{fig:manifold}~(a)] and the angle $\varphi$ [purple cross for
$\varphi=0$, orange cross for $\varphi=\pi$ in Fig.~\ref{fig:manifold}~(a)] with the $\Omega$-axis in the $\Gamma = \mm{const.}$
plane. 

When $\gamma_\mm{circ}(\varepsilon)$ is centered at $\pv_\mm{center}$, the function $\chi$ has the simple form
\begin{equation}
    \chi(\varepsilon)= \pi \Theta \left[ \frac{1-d}{2}+ \left(\varepsilon - \frac{1}{2} +\frac{\varphi}{2\pi}\right) \right],
    \label{eq:ChiEpsilon}
\end{equation}
where $\Theta (x)$ is the Heaviside step function (with the convention $\Theta (0)=1$) and the parameter $d$ accounts for the loop
orientation ($\pm 1$ for clockwise/counter-clockwise). For the rest of this article, and unless otherwise specified, we consider
the closed contour $\gamma_\mm{circ}(\varepsilon)$ centered at $\pv_\mm{center}$ with basepoint $\varphi = 0$ or $\varphi = \pi$.

Using Eqs.~\eqref{eq:cont_lambda_static} and \eqref{eq:ChiEpsilon}, we find that $\lambda_\pm [\pv (\varepsilon)]$ trace a
continuous eigenvalue braid that obeys
\begin{equation}
    \lambda_\pm[\pv(0)] \to -\lambda_\pm[\pv(1)] = \lambda_\mp [\pv(0)].
    \label{eq:braiding}
\end{equation}
While Eq.~\eqref{eq:braiding} is valid for any basepoint $\varphi$, the value of $\lambda_\pm[\pv(0)]$ (end points of the braid)
depends explicitly on the choice of basepoint $\varphi$ (braids generated from loops with different basepoints are not isotopic
since the isotopy of braids requires the end point to be fixed). We show this in Fig.~\ref{fig:manifold} (c) where the braids
traced for $\varphi=0$ and $\varphi=\pi$ are compared. 

The mapping of the initial to the final eigenvalues is described by the permutation $\sigma_1 (\psi_\pm) = \psi_\mp$ [see
Eq.~\eqref{eq:EigValPermutation}]. The corresponding topological operation is given by
\begin{equation}
    \hat{P}_{\sigma_1} = \ketbra{\psi_+ [\pv(0)]}{\psi_- [\pv(0)]} + \ketbra{\psi_- [\pv(0)]}{\psi_+ [\pv(0)]},
    \label{eq:sigma_x}
\end{equation}
which follows from Eq.~\eqref{eq:EigVecPermutation}. Since the mapping described by Eq.~\eqref{eq:braiding} is
independent of the choice of contour (including the basepoint), $\sigma^\ell_n=\prod_{k=1}^\ell \sigma_n$ describes the
eigenvalue permutation for any loop that winds $\ell$ times around $B_2$. 

The topological operation defined in Eq.~\eqref{eq:sigma_x} depends on the choice of basepoint $\varphi$, since the
eigenmodes explicitly depend on $\varphi$ [see Eqs.~\eqref{eq:StaticEigVecs} and \eqref{eq:Num_delta_Num_omega}]. This implies
that $\hat{P}_{\sigma_1}(\varphi=0)\neq\hat{P}_{\sigma_1}(\varphi=\pi)$.

\subsection{Non-Hermitian Systems and Violation of the Adiabatic Theorem} 
\label{subsec:dynamics}

The easiest way to implement the operation defined in Eq.~\eqref{eq:sigma_x} would be to adiabatically evolve $\hH_\mm{sym}$ along
$\gamma_\mm{circ} (\varepsilon)$. As discussed in Sec.~\ref{subsec:prelim_dynamics} and extensively in the
literature~\cite{xu2016,nenciu1980,Uzdin2011,Berry2011}, this is impossible for non-Hermitian systems. However, for completeness
and to have a baseline to compare against non-Hermitian STAs, we briefly discuss below the dynamics generated by $\hH_\mm{sym}$
when $\pv$ varies slowly in time. 

To this end, we turn $\varepsilon$ into a smooth, monotonic time-varying function defined by
\begin{equation} 
\label{eq:epsilon_time}
    \varepsilon(t)=
    d \left[6 \left(\frac{t}{t_0} \right)^5 -15 \left(\frac{t}{t_0} \right)^4  +10 \left(\frac{t}{t_0} \right)^3 \right],
\end{equation}
which fulfills $\dot{\varepsilon} (0) = \dot{\varepsilon} (t_0) = \ddot{\varepsilon} (0) = \ddot{\varepsilon} (t_0) =0$ (where
$\ddot{x}=\ud^2x/\ud t^2$). This function varies from $0$ to $1$ for $t \in [0, t_0]$, with $t_0$ being the time needed to trace
the loop $\gamma_\mm{circ} (\varepsilon)$. We refer to the time-generated loop as the control loop, and use the
dimensionless quantity $\Gamma_0 t_0$ to refer to the loop time duration.

Expressing $\hH_\mm{sym} (t)$ [see Eq.~\eqref{eq:H_sym_static}] in the frame of instantaneous eigenmodes via
$\hS_\mm{ad}(t)$ [see Eqs.~\eqref{eq:GenericTimeDepH} and \eqref{eq:S_ad_holo}], we obtain the adiabatic NH Hamiltonian 
\begin{equation} 
\label{eq:H_ad_dyn}
    H_{\mm{ad}}(t)=\lambda(t)\hat{\sigma}_{z,\mm{ad}}- \frac{1}{2}\dot\theta(t)\hat{\sigma}_{y,\mm{ad}},
\end{equation}
which we use to solve for the flow $\hat\Phi_\mm{ad} (t)$ [see Eq.~\eqref{eq:SchrodingerEq}]. Transition
probabilities between instantaneous eigenmodes are then given by
\begin{equation}
    P_{i,j}(t) = \frac{\left|\bra{\psi_j} \hat\Phi_\mm{ad}(t) \ket{\psi_i}\right|^2}{\sum_{j} \left| \bra{\psi_j} \hat\Phi_\mm{ad} (t)\ket{\psi_i} \right|^2},
    \label{eq:Prob}
\end{equation}
with $i,j \in \{+,-\}$.

\begin{figure}[t] 
    \centering
    \includegraphics[width=\columnwidth]{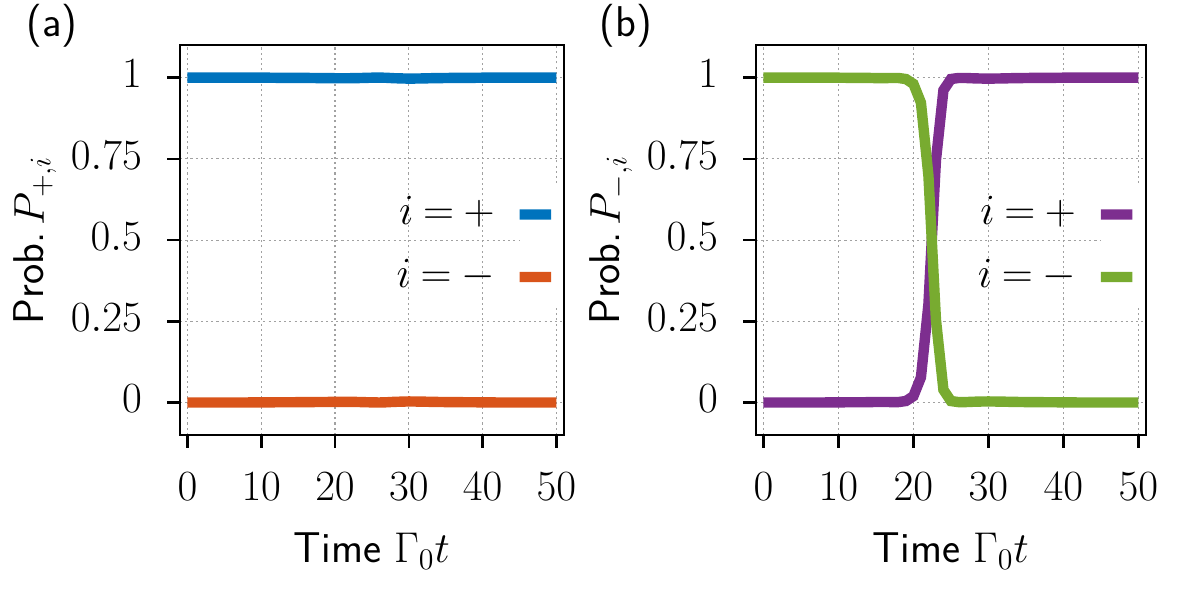}
    \caption{Non-reciprocal dynamics with a two-mode system. Implementing a clockwise control loop around an EP leads to (a)
    $P_{+,+} (t_0) = 1$ (blue trace), $P_{-,+} = 0$ (orange trace) and (b) $P_{-,+} (t_0) = 1$ (purple trace), $P_{-,-} (t_0) = 0$
    (green trace).
    }
    \label{fig:gain_loss}
\end{figure}

We plot in Fig.~\ref{fig:gain_loss} $P_{i,j} (t)$ as a function of $\Gamma_0 t$ for $\Delta_0=\Omega_0=\Gamma_0/2, \Gamma_0 t_0
=50, \varphi=0, d=1$. When the system is initialized in the least damped mode ($\ket{\psi_+}$), we find as expected that
$P_{+,+}(t) \simeq 1$ and $P_{+,-}(t) \simeq 0$ (blue and orange traces in Fig.~\ref{fig:gain_loss}(a), respectively) for all
times. This indicates that $\ket{\psi_+}$ evolves adiabatically and $\lambda_+ (t)$ traces the expected strand of the braid
associated with $\gamma_\mm{circ} (\varepsilon)$. However, when the system is initialized in the most damped mode
($\ket{\psi_-}$), we have $P_{-,-}(t) \neq 1$ and $P_{-,+}(t) \neq 0$ for all times (green and purple traces in
Fig.~\ref{fig:gain_loss} (b), respectively), indicating that $\ket{\psi_-}$ is not evolving adiabatically; the small non-adiabatic
transitions from $\ket{\psi_-} \to \ket{\psi_+}$ get exponentially enhanced due to amplitude amplification of the least damped
mode~\cite{Wang2018,Nenciu1992,Moiseyev2011}. As a consequence, the strand of the braid associated to $\lambda_- (t)$ cannot be
traced and the topological operation defined in Eq.~\eqref{eq:sigma_x} cannot be generated.

\subsection{Near Ideal Permutation Operation via Near-Coherent Evolution} 
\label{subsec:phi_eq_pi}

\begin{figure}[t] 
    \centering
    \includegraphics[width=\columnwidth]{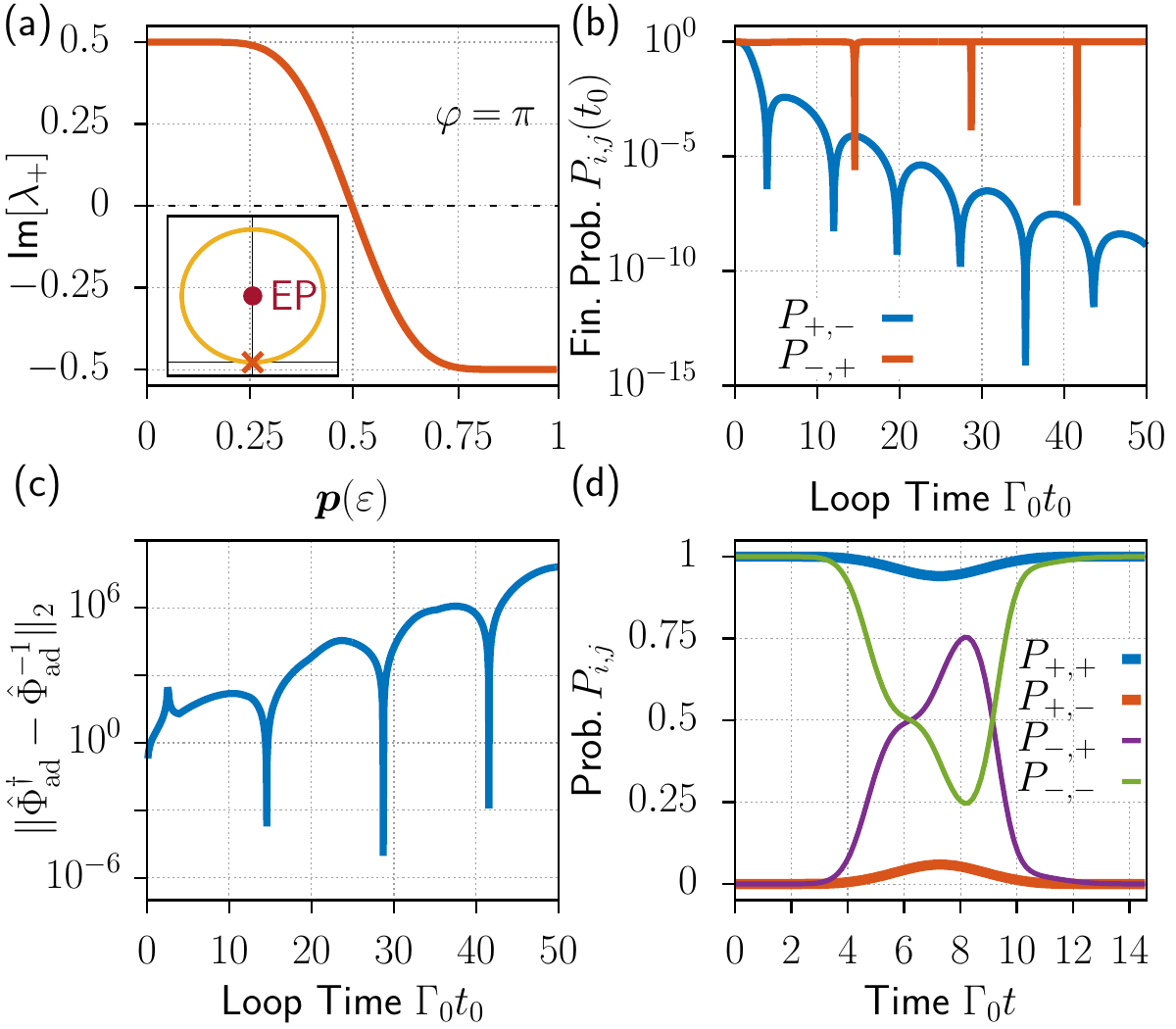}
    \caption{Dynamics with contours yielding near ideal topological operations. (a) Choosing $\varphi=\pi$ as a basepoint for
        $\gamma_\mm{circ} (\varepsilon)$  yields a spectrum with an antisymmetric imaginary part that balances on average gain and
        loss. (b) Akin to coherent evolution there exists special loop times for which the final transitions probabilities
        $P_{+,-} (t_0)$ and $P_{-,+} (t_0)$ vanish simultaneously. (c) For these special loop times, the flow associated with the NH
        Hamiltonian at $t=t_0$ is quasi-unitary, i.e., $\hat{\Phi}^\dagger (t_0) \approx \hat{\Phi}^{-1} (t_0)$. (d) Transition
        probabilities as a function of $\Gamma_0 t$ for the special loop time $\Gamma_0 t_0 = 14.5$.
    }
    \label{fig:phieqpi} 
\end{figure}

Among the (infinite) possibilities for a basepoint along $\gamma_\mm{circ} (\varepsilon)$, there is a special choice corresponding
to $\varphi=\pi$ that leads to $\mm{Im}[\lambda_\pm [\pv(\varepsilon)]$ being an anti-symmetric function with respect to
$\varepsilon =1/2$ [see Fig.~\ref{fig:phieqpi}(a)]. This leads to $\hat\Phi_\mm{ad} (t_0)$ being ``quasi'' unitary for certain
loop times $\Gamma_0 t_0$ since the imaginary part of the eigenvalues average out to zero during the evolution. We, thus,
expect to observe interference phenomena akin of what has been observed in non-Hermitian PT-symmetric
systems~\cite{Klauck2019,Ornigotti2014,Scheel2018,Guo2009,Ghosh2021}. 

We plot in Fig.~\ref{fig:phieqpi}(b) the non-adiabatic transitions probabilities $P_{+,-} (t_0)$ (blue trace) and $P_{-,+} (t_0)$
(orange trace) evaluated at $t=t_0$ as a function of the loop time $\Gamma_0 t_0$. Our results show that for $\Gamma_0 t_0 > 1$
the non-adiabatic transition probability $P_{+,-} (t_0)$ almost vanishes. This necessarily implies that if the system is
initialized in the least damped eigenmode ($\ket{\psi_+}$) at $t=0$, it will end up in the same eigenmode at $t=t_0$. The
situation is quite different when the system is initialized in the most damped mode ($\ket{\psi_-}$). For almost all values of
$\Gamma_0 t_0$ the non-adiabatic transition probability is maximal $P_{-,+} (t_0) \simeq 1$, but there are special values of
$\Gamma_0 t_0$ at which $P_{-,+} (t_0)$ vanishes. For those special times, the system ends up in the most damped eigenmode, if it
is initially prepared in that eigenmode. 

To verify our interpretation of the dynamics, we check that the flow $\hat\Phi_\mm{ad} (t_0)$ is ``quasi'' unitary at the special
values of $\Gamma_0 t_0$. In Fig.~\ref{fig:phieqpi}(c), we plot the spectral norm $||\hat\Phi^\dagger_\mm{ad}(t_0) -
\hat\Phi_\mm{ad}^{-1} (t_0)||_2$ as a function of $\Gamma_0 t_0$. When the spectral norm is $0$, we have $\hat\Phi_\mm{ad}^\dagger
(t_0) = \hat\Phi_\mm{ad} (t_0)$, which is the defining property of unitary operators. 

In Fig.~\ref{fig:phieqpi}(d), we plot the transition probabilities $P_{+,+} (t)$ (thick blue trace),  $P_{+,-} (t)$ (thick orange
trace), $P_{-,+} (t)$ (thin purple trace), and $P_{-,-} (t)$ (thin green trace) as a function of $\Gamma_0 t$ for the special
control loop time $\Gamma_0 t_0 =14.5$. As one can observe the evolution of the eigenmodes is not adiabatic, but akin to a
coherent evolution where destructive interference brings the system back to the desired eigenmode at $t=t_0$ after populating the
other eigenmode at intermediate times. This also corresponds to the physical principle behind STAs, which only emulate adiabatic
dynamics at the final time. 

The difficulty in using these types of control loops to generate topological operation lies in achieving the exact timing at which
$P_{-,+} (t_0)$ vanishes; any small deviations from these special loop times will result in $P_{-,+} (t_0) > 0$ [see orange trace
Fig.~\ref{fig:phieqpi}(b)]. The idea of using a control loop that yields a spectrum with an anti-symmetric imaginary part with
respect to $\varepsilon/2$ to emulate adiabatic dynamics has been previously reported~\cite{Arkhipov2024,wu2024,Wu2024_2}. 

\subsection{Transitionless Driving} 
\label{sec:TD}

We are now ready to discuss the main results of this manuscript, the construction of non-Hermitian STAs. We start by deriving  the
non-Hermitian equivalent of the well-known STA coined Transitionless Driving (TD)~\cite{demirplak2003, berry2009}, which is
designed to suppress non-adiabatic transitions between the instantaneous eigenstates at all times. The dressing transformation
[see Eq.~\eqref{eq:GenSdrSTA}] leading to TD is the trivial dressing
\begin{equation} 
    \hS_\mm{TD}(t)=\mathbbm{1},
\label{eq:TD_operator}
\end{equation}
which guarantees that the dressed states $\ket{\phi_j}$ [see Eq.~\eqref{eq:GenSdrSTA}] correspond at all times to the
instantaneous eigenmodes $\ket{\psi_j}$ of $\hH_\mm{sym}$ [see Eq.~\eqref{eq:GenSAd}].

Using Eq.~\eqref{eq:H_dr_condition}, one readily finds the non-Hermitian control Hamiltonian associated to TD. In the frame of
instantaneous eigenmodes of $\hH_\mm{ad}(t)$, we have
\begin{equation} 
    \hW_\mm{ad,TD} (t) = +i \hS_\mm{ad}^{-1} (t) \rd_t\hS_\mm{ad} (t) = \frac{1}{2}\dot{\theta} (t) \hat{\sigma}_{y,\mm{ad}},
    \label{eq:W_TD_ad}
\end{equation}
where $\theta (t) = \theta\{\gamma_\mm{circ}[\varepsilon(t)]\}$ with $\theta(\pv)$, $\gamma_\mm{circ} (\varepsilon)$, and
$\varepsilon(t)$ defined in Eqs.~\eqref{eq:cont_theta_static}, \eqref{eq:Num_delta_Num_omega}, and \eqref{eq:epsilon_time},
respectively. Like in the Hermitian case, $\hW_\mm{ad,TD} (t)$ is simply the opposite of the non-inertial term.

We stress once more the importance of constructing change of frame operators that are holomorphic to obtain the correct
non-inertial term [see Sec.~\ref{subsec:STAs}]. This includes the change of frame operator that diagonalizes the non-Hermitian
Hamiltonian~[see Eq.~\eqref{eq:H_sym_static}] as well as the dressing transformation. If any of these are not holomorphic, it will
lead to unphysical $\hW$'s. 

For the two-mode example considered here, this boils down to either defining the eigenvalues $\lambda (\pv)$ [see
Eqs.~\eqref{eq:cont_lambda_static} and \eqref{eq:StaticEigVecs}] or the pseudo-rotation angle $\theta (\pv)$ to be a holomorphic
function of $\pv$ [see Eqs.~\eqref{eq:S_ad_holo} and \eqref{eq:cont_theta_static}]. However, for this specific example, since we
have $\dot{\chi}\{\gamma_\mm{circ} [\varepsilon (t)]\} =0$ [see Eqs.~\eqref{eq:cont_theta_static}, \eqref{eq:ChiEpsilon}, and
\eqref{eq:epsilon_time}], using a non-holomorphic version of $\lambda (\pv)$ or $\theta (\pv)$ will also lead to the correct
control operator [see Eq.~\eqref{eq:W_TD_ad}]. We discuss in Sec.~\ref{sec:SATDCtrlW} how failing to define the change of
frame operator to be holomorphic generally leads to unphysical situations when considering higher-order STAs.

\begin{figure}[t] 
    \includegraphics[width=\columnwidth]{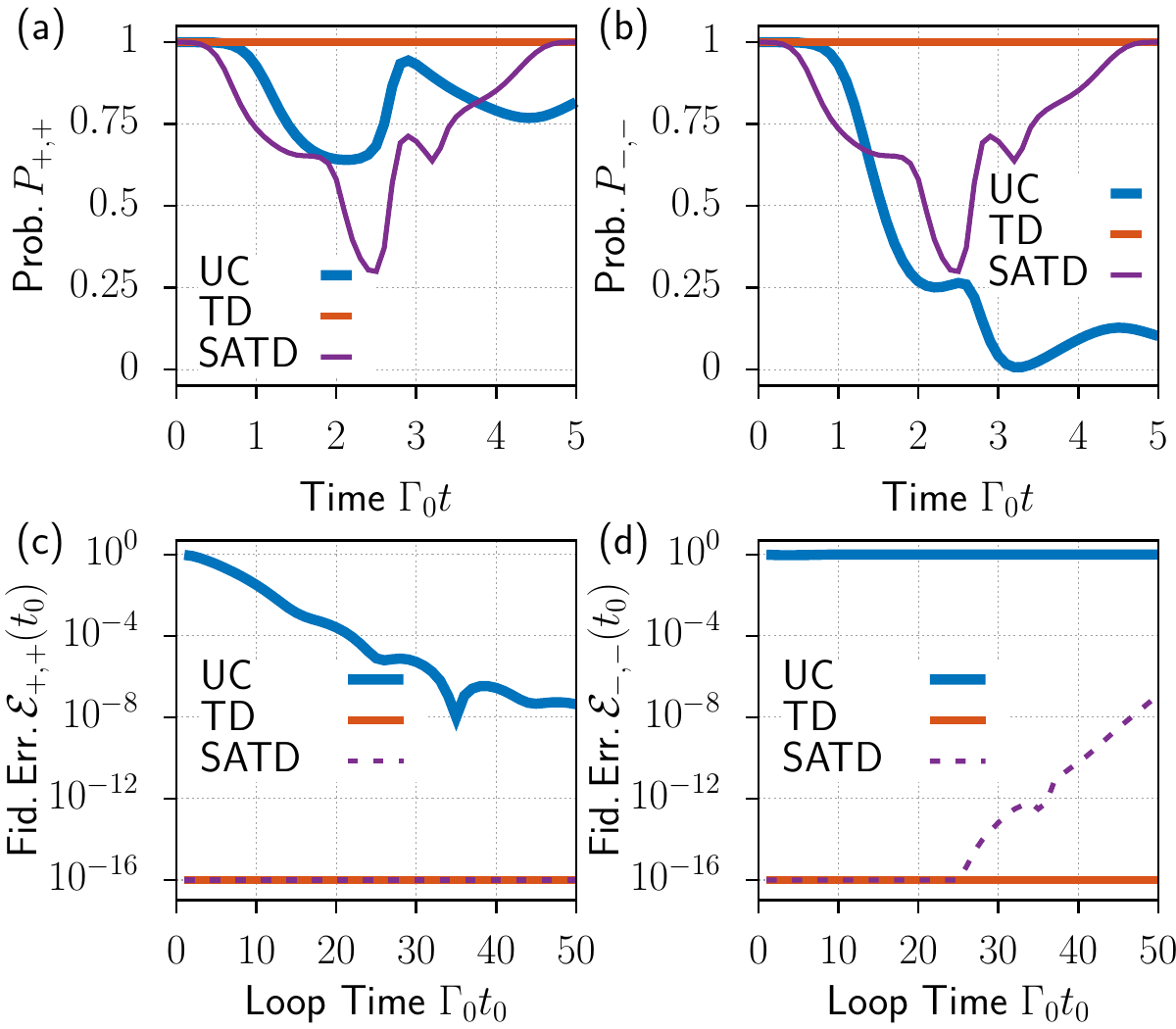}
    \caption{Comparison between uncorrected (UC, thick blue trace), TD (orange trace), and SATD (thin purple trace) control protocols. (a) Transition probability
        $P_{+,+} (t)$ and (b) $P_{-,-} (t)$ for $\Gamma_0 t_0 = 5$. (c) State dependent fidelity error $\mathcal{E}_{+,+} (t_0)$
        and (d)  $\mathcal{E}_{-,-} (t_0)$ as a function of $\Gamma_0 t_0$. TD traces the ideal braid at
        all times (see Sec.~\ref{sec:TD}), while SATD 
        guarantees that the end points of the generated braids coincide with those of the ideal case (see Sec.~\ref{sec:SATD}).}
    \label{fig:corr_err_combo}
\end{figure}

Like in the Hermitian case, the modified non-Hermitian Hamiltonian $\hH_\mm{sym,TD} (t) =\hH_\mm{sym} (t) + \hW_\mm{sym,TD} (t)$,
with $\hW_\mm{sym,TD} (t) = \hS_\mm{ad}(t) \hW_\mm{ad,TD} (t) \hS_\mm{ad}^{-1} (t)$, emulates the expected ideal adiabatic
dynamics. We compare in Fig.~\ref{fig:corr_err_combo}(a) [Fig.~\ref{fig:corr_err_combo}(b)] the transition probabilities $P_{+,+}
(t)$ [$P_{-,-} (t)]$ obtained with an uncorrected protocol (thick blue trace) against the TD protocol (orange trace).  As expected
with TD, we have $P_{+,+} (t) = P_{-,-} (t) = 1$ which indicates that the system remains at all times in the initial eigenmode.

We also compare in Figs.~\ref{fig:corr_err_combo}(c) and (d) the state-dependent fidelity error
\begin{equation} 
    \mathcal{E}_{i,i}(t_0) = 1-P_{i,i}(t_0)
\label{eq:fidelity_error}
\end{equation}
as a function of the control loop time $\Gamma_0 t_0$ for $i \in \{+,-\}$ between the uncorrected protocol (thick blue trace) and TD
(orange trace). Our results show that TD behaves as in the Hermitian case, meaning one can view
$\hW_\mm{sym,TD} (t)$ as a family of control operators labeled by the parameter $t_0$. As we discuss below, and in contrast to the
Hermitian case, this is in general not true for higher-order non-Hermitian STAs. 

The numerical simulations were done using $\Delta_0=\frac{1}{2}, \Omega_0=\frac{1}{6}, \varphi = -\frac{\pi}{8}, d=1$, and
$\Gamma_0 t_0=5$ for Figs.~\ref{fig:corr_err_combo}(a) and (b), and $\Delta_0=\Omega_0=\Gamma_0/2, \varphi=0$, and $d=1$ for
Figs.~\ref{fig:corr_err_combo}(c) and (d). 

\subsubsection{Transitionless driving dynamics} 
\label{sec:TD_dynamics}

For the Hermitian case, the TD correction modifies the \emph{coherent} dynamics of the system such that when the system is
initialized in one of the old eigenmodes [an eigenmode of $\hH_\mm{sym}(t)$ as opposed to an eigenmode of
$\hH_\mm{sym,TD}(t)$], destructive interference prevents transitions at all times to the other old eigenmode.

For non-Hermitian systems, this interpretation cannot hold since we are dealing with dissipative dynamics. The generalization
would be that the non-Hermitian version of TD modifies the \emph{incoherent} dynamics such that the effective decay rate of an old
eigenmode corresponds to the ``ideal'' decay rate associated to the adiabatic approximation of $\hH_\mm{sym}(t)$.  This implies,
like in the Hermitian case, that $\hH_\mm{sym,TD} (t)$ generates a dynamics that mimics the idealized adiabatic dynamics of
$\hH_\mm{sym}$.  There is, however, another possible explanation for what the non-Hermitian TD correction does. It modifies the
control loops such that they wind around an exceptional point of the modified spectrum [the spectrum of $\hH_\mm{sym,TD} (t)$] in
such a way that the generated dynamics happens to be close to the ideal adiabatic dynamics of $\hH_\mm{sym}(t)$ (similar to having
control loops that lead to the imaginary part of the spectrum to be anti-symmetric, see Sec.~\ref{subsec:phi_eq_pi}).

To distinguish between these two scenarios, we checked if the modified control loops wind around an EP associated to
$\hH_\mm{sym,TD}(t)$ (see Appendix~\ref{sec:app_eigeval_braid}). 

\begin{figure}[t] 
    \includegraphics[width=\columnwidth]{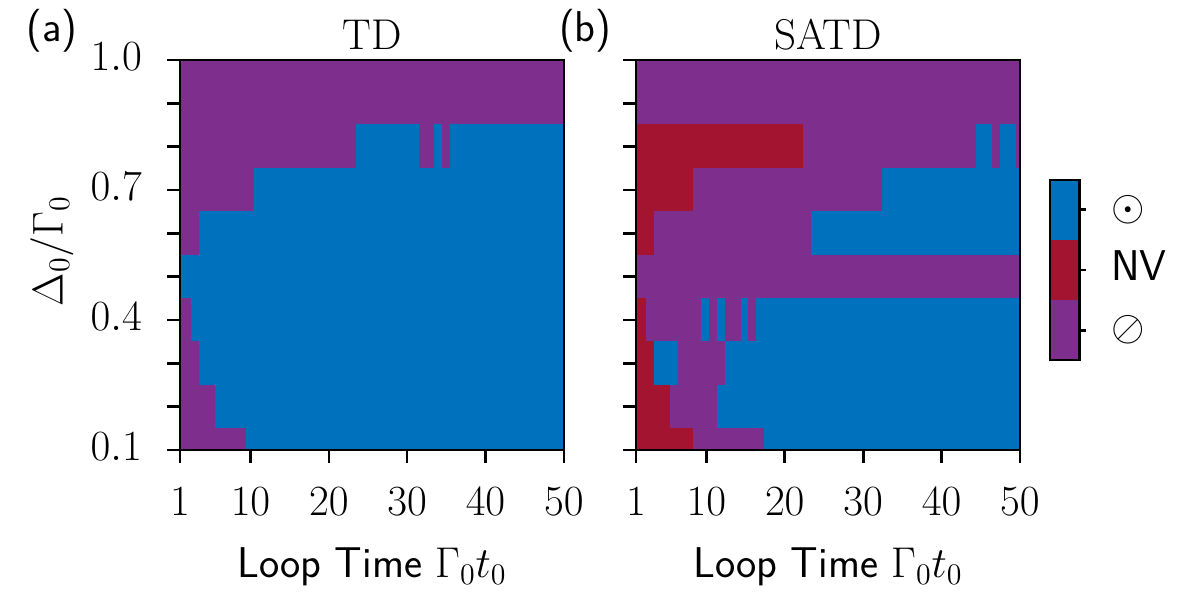}
    \caption{STAs validity. Exceptional point encircling check [see Eq.~\eqref{eq:ep_encircle_braiding_conditons}] for (a) TD correction [see
    Sec.~\ref{sec:TD}] and (b) SATD correction [see Sec.~\ref{sec:SATD}]. We find that TD and SATD ensure eigenvalue
    braiding for all (applicable) parameter sets, but the EP associated to $\hH_\mm{sym,STA}$ is not necessarily encircled. The labeling
    scheme is as follows: $\odot$ - EP encircled, $\oslash$ - EP not encircled, NV - not a valid STA. Braiding of the eigenvalues
    associated to $\hH_\mm{sym,mod}$ is observed for all (applicable) parameter sets.}
    \label{fig:STA_SLDL}
\end{figure}

We plot in Fig.~\ref{fig:STA_SLDL}(a) as a function of loop time $\Gamma_0 t_0$ and radius $\Delta_0/\Gamma_0$ [see
Eq.~\eqref{eq:Num_delta_Num_omega}] a map showing when the TD modified control loop winds around an EP associated to the spectrum
of $\hH_\mm{sym,TD} (t)$ (blue) or does not (purple). We check this for $1\leq\Gamma_0 t_0\leq 50,0.1\leq\Delta_0 t_0\leq 1,
\varphi=0,\Omega_0=\Gamma_0/2$, and $d=1$. Our result show that the modified control loop does not necessarily wind around an EP
of the modified spectrum, but as discussed earlier, the dynamics always yields an eigenvalue braid.  This indicates that the
non-Hermitian version of TD, similar to the Hermitian case, emulates the ideal adiabatic dynamics. 

\subsubsection{Choice of initial control loop}
\label{sec:TDChoiceInitLoop}

In Sec.~\ref{subsec:phi_eq_pi} we showed that a different choice of basepoint along the same closed contour can result in control
loops that generate a dynamics yielding the desired topological operation [see Eq.~\eqref{eq:sigma_x}] without tracing the ideal
eigenvalue braid. Since this particular choice of control loop can lead to the desired dynamics without applying a correction, one
might presume that it will lead to a more energy-effective TD correction since there is ``less'' to correct.

\begin{figure}[t] 
    \includegraphics[width=\columnwidth]{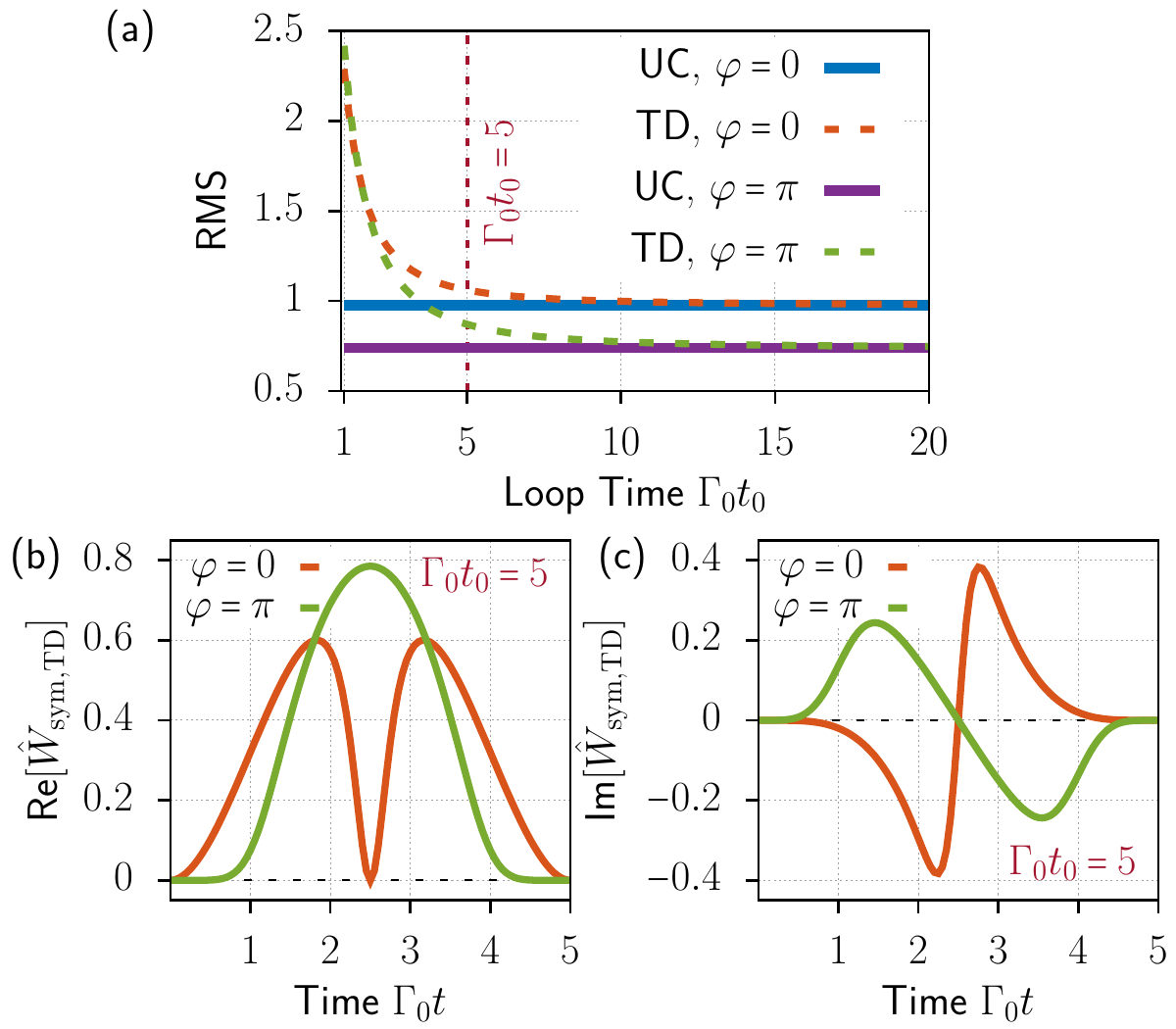}
    \caption{RMS amplitude for TD correction. (a) Comparison of RMS [see Eq.~\eqref{eq:RMS}] amplitudes between uncorrected case (UC,
        solid lines) and TD (dashed lines) for $\varphi=0$ (blue and orange traces) and $\varphi=\pi$ (purple and green). Example
        of (b) real and (c) imaginary parts of the TD correction $\hW_\mm{TD}$ for $\Gamma_0 t_0 = 14.5$. Leveraging the quasi-coherent
        regime of the uncorrected dynamics does yield TD fields with reduced energy (see main text for more details).
    }
    \label{fig:TD_phi_eq_pi}
\end{figure}

We plot in Fig.~\ref{fig:TD_phi_eq_pi}(a) the root mean square (RMS) amplitude of the fields used to implement a specific dynamics
as a function of control loop time $\Gamma_0 t_0$. Since the control fields can be complex, we define the RMS amplitude to be
\begin{equation} 
    \mm{RMS} = \sqrt{\frac{1}{t_0}\int_0^{t_0} \di{t} \sum_{i=x,y,z} \abs{f_i (t)}^2}.
\label{eq:RMS}
\end{equation}
where $f_i (t)$ is the coefficient of the decomposition of $\hH(t)$ associated to the Pauli operator $hat{\sigma}_i$ with $i\in
\{x,y,z\}$. We show this for the uncorrected fields (UC) associated to the initial control loop [see
Eq.~\eqref{eq:Num_delta_Num_omega}], and the TD correction $\hW_\mm{sym,TD}$ for $\varphi=0$ and $\varphi=\pi$. We used
$\Delta_0=\Omega_0=\Gamma_0/2,$ and $d=1$.

Our results indicate that there is no significant energy reduction in implementing TD when choosing an initial control loop with
basepoint $\varphi=\pi$; more precisely for $\Gamma_0 t_0 \gtrsim 5$ the RMS field amplitude of $\hH_\mm{sym}(t)$ (solid traces)
and $\hH_\mm{sym,TD} (t)$ (dashed traces) is nearly the same for both $\varphi=0$ (solid blue and dashed orange traces) and
$\varphi=\pi$ (solid purple and dashed green traces). 

We further check that there is no significant difference in the control field amplitudes for the special loop times that generate
``STA-like'' dynamics when $\varphi=\pi$ [see Fig.~\ref{fig:phieqpi}].  We show this behavior in Figs.~\ref{fig:TD_phi_eq_pi}(b)
and \ref{fig:TD_phi_eq_pi}(c) where we plot the real and imaginary parts, respectively, of the matrix elements of
$\hW_\mm{sym,TD}$ for $\varphi=0$ (orange trace) and $\varphi=\pi$ (green trace) for $\Gamma_0 t_0\approx14.5$.

This is to be expected since the TD correction is designed to cancel non-adiabatic transitions between the eigenmodes of
$\hH_\mm{sym}$ at all times. From a resource-cost point of view, TD can therefore not benefit from the dynamics happening at
special values of $\Gamma_0 t_0$ and $\varphi=\pi$, since there are still non-adiabatic transitions that need to be canceled at
intermediate times.

\subsection{Super Adiabatic Transitionless Driving} 
\label{sec:SATD}

Despite its effectiveness, the TD correction may be considered too demanding, as it often requires considerable resources to be
implemented~\cite{Bason2012}, e.g., for the two-mode model considered here one must be able to couple to $\hat{\sigma}_y$ and tune
the interaction. To overcome this limitation, an STA coined superadiabatic transitionless driving (SATD) was
introduced~\cite{demirplak2008,baksic2016}. SATD is known to yield correction fields that are compatible with the constraints of a
two-level system and a lambda system~\cite{Zhou2017,baksic2016}.

Fundamentally, SATD is a special case of the dressed state approach to STAs in which the dressing is chosen to diagonalize the
adiabatic Hamiltonian (which includes the non-inertial term that couples the instantaneous eigenmodes [see
Eq.~\eqref{eq:H_ad_dyn}]). 

For Hermitian systems, independently of the protocol time $t_0$, SATD allows one to engineer a modified Hamiltonian
$\hH_\mm{sym,SATD} (t)$ that evolves a set of chosen eigenstates $\ket{\psi_i (t)}$ of $\hH(t)$ such that $\ket{\psi_i (0)}$ is
mapped to $\ket{\psi_i (t_0)}$ without enforcing the system to occupy $\ket{\psi_i (t)}$ at all intermediate times.  For
non-Hermitian systems, as we discuss below, SATD does not yield for all values of $t_0$ a modified Hamiltonian that generates the
desired mapping. The SATD solutions cannot, therefore, be considered as a family of solutions labeled by the protocol time $t_0$. 

\subsubsection{SATD Dressing Transformation}
\label{sec:SATD_DT}

For the two-mode system described by Eq.~\eqref{eq:H_sym_static}, the SATD dressing operator, which coincides with the change of
frame operator that diagonalizes $\hH_\mm{ad} (t)$, is given by 
\begin{equation} 
    \hS_{\mm{SATD}} (t) = \exp \left[-\frac{i}{2} \mu (t) \hat{\sigma}_x\right],
    \label{eq:S_dr_satd}
\end{equation} 
where the holomorphic dressing angle is
\begin{equation} 
    \begin{aligned}
        \mu (t) &=  \mu_\mm{nat} (t) + \chi_\mu (t)\\
                &= - \arctan\left[\frac{1}{2} \frac{\dot{\theta} (t)}{\lambda (t)} \right] + \chi_\mu (t).
    \end{aligned}
    %\mu (t) = - dd\arctan\left[ \frac{\frac{1}{2}\dot{\theta} (t)}{\lambda (t)} \right] \pm n\pi,
    \label{eq:mu_dr_holo}
\end{equation}
Here, $\lambda (t) $ and $\theta (t)$ are given by Eqs.~\eqref{eq:cont_lambda_static} and \eqref{eq:cont_theta_static},
respectively, and the time dependence is acquired through a choice of control loop $\gamma [\varepsilon (t)]$, i.e., $\mu (t) =
\mu\{\gamma[\varepsilon (t)]\}$. The function $\chi_\mu (t)$ is defined by
\begin{equation}
    \chi_\mu (t) = \pi \sum_s \Theta (t-t_s),
    \label{eq:SATDchi}
\end{equation}
with $\Theta (x)$ defined below Eq.~\eqref{eq:ChiEpsilon} and $t_s$ corresponds to the times at which the control loop ``crosses''
the branch cut associated to $\mu_\mm{nat} (t)$. Geometrically, one can view $\chi_\mu (t)$ as a function that glues together the
leaves of $\mu_\mm{nat} (t)$ to make up the Riemann surface, on which $\mu (t)$ is holomorphic.

Constructing the holomorphic dressing angle from the natural dressing angle does not necessarily guarantee that the boundary
properties are conserved. Equation~\eqref{eq:mu_dr_holo} shows that even if one has initially engineered a control loop that
fulfills $\mu_\mm{nat} (0) = \mu_\mm{nat}(t_0) = 0$, we have $\mu (t_0) = n \pi$, where $n$ is the number of times the
branch cut associated to $\mu (t)$ is crossed. Thus, $\hS_{\mm{SATD}} (t_0)$ is not necessarily the identity transformation as
required (see Sec.~\ref{subsec:STAs}) since
\begin{equation} 
    \hS_\mm{SATD} (t_0) = \exp{\left[-\frac{i}{2}(n \pi) \hat{\sigma}_x\right]}=(-i)^n\hat{\sigma}_x^n.
    \label{eq:SATD_breaking}
\end{equation}
If $n$ is an even integer then Eq.~\eqref{eq:SATD_breaking} reduces to $(i)^n \mathbbm{1}$ and Eq.~\eqref{eq:STA_condition} is
fulfilled up to an (irrelevant) global phase.  However, if $n$ is an odd integer, then Eq.~\eqref{eq:SATD_breaking} reduces to
$(i)^n \hat{\sigma}_x$, which violates Eq.~\eqref{eq:STA_condition} and immediately shows that in this case one cannot obtain a
valid STA. 

\begin{figure}[t] 
    \includegraphics[width=\columnwidth]{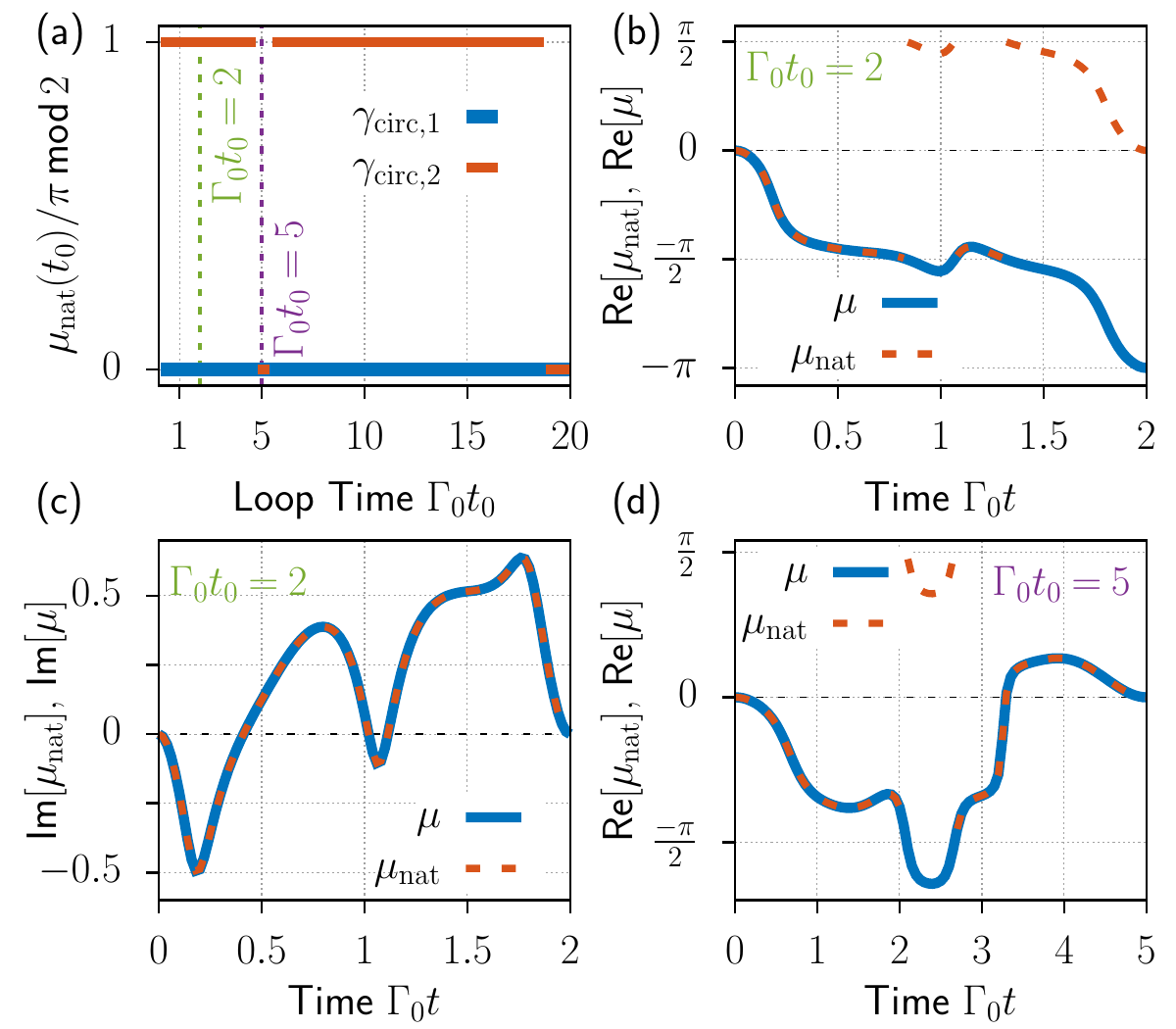}
    \caption{Determining the validity of SATD. (a) SATD is not a valid STA over the whole range $\Gamma_0 t_0 = [0,25]$ 
        for $\gamma_{\mm{circ},2}$ (see main text for details)
        while it is a valid STA for $\gamma_{\mm{circ},1}$ over the same range $\Gamma_0 t_0$. 
        Real (b) and imaginary (c) parts of $\mu (t)$ and $\mu_\mm{nat} (t)$ [see Eq.~\eqref{eq:mu_dr_holo}] obtained with
        $\gamma_{\mm{circ},2}$ leading to an invalid STA
        ($\Gamma_0 t_0 =2$). (d) Same as (b) but using $\gamma_{\mm{circ},1}$, which leads to a valid STA ($\Gamma_0 t_0 =5$).
    }
    \label{fig:mu_islands}
\end{figure}

For a choice of control loop $\gamma[\varepsilon (t)]$, there are several parameters that can control how many times the branch
cut associated to $\mu_\mm{nat} (t)$, including the loop time $\Gamma_0 t_0$. Thus, and in stark contrast to the Hermitian case,
the non-Hermitian version of SATD cannot be seen as a family of solutions labeled by the control loop time $\Gamma_0 t_0$. We
illustrate this behavior in Fig.~\ref{fig:mu_islands}(a) where we plot the value of $\mu(t_0)/\pi \mm{Mod} 2$ as a function of
$\Gamma_0 t_0$ for two different circular contours $\gamma_{\mm{circ},1} [\varepsilon(t)]$ (blue trace) and
$\gamma_{\mm{circ},2}[\varepsilon(t)]$ (orange trace) [see Eq.~\eqref{eq:epsilon_time}]. 

We choose $\Delta_0=\Omega_0=\Gamma_0/2, \varphi=0, d=1$ for $\gamma_{\mm{circ},1}$, and $\Delta_0=\Gamma_0/2,
\Omega_0=\Gamma_0/6, \varphi = -\pi/8, d=1$ for $\gamma_{\mm{circ},2}$. The dressing angle obtained from $\gamma_{\mm{circ},1}
[\varepsilon(t)]$ always yields a valid SATD protocol, i.e., the dressing transformation obeys conditions (1)-(3) of the dressed
state approach to non-Hermitian STAs (see Sec.~\ref{subsec:STAs}), while the dressing angle obtained from $\gamma_{\mm{circ},2}
[\varepsilon(t)]$ only leads to a valid SATD protocol for certain values of $\Gamma_0 t_0$.

We plot in Figs.~\ref{fig:mu_islands}(b) and (c) the real and imaginary parts of $\mu_\mm{nat} (t)$ (dashed orange trace) and $\mu
(t)$ (solid blue trace) [see Eq.~\eqref{eq:mu_dr_holo}] obtained with $\gamma_{\mm{circ},2} [\varepsilon(t)]$ as a function of time
for $\Gamma_0 t_0 = 2$ [corresponding to the green dashed line in Fig.~\ref{fig:mu_islands}(a)]. Turning the natural dressing
angle into its holomorphic counterpart leads to a dressing angle that does not vanish at $t=t_0$ and, thus, cannot be used to
define a valid SATD dressing transformation. In contrast, for $\Gamma_0 t_0 = 5$ [corresponding to the purple dashed line in
Fig.~\ref{fig:mu_islands}(a)], one obtains a holomorphic dressing angle that vanishes at the end points and leads to a valid SATD
protocol. This is illustrated in Fig.~\ref{fig:mu_islands}(d) where we plot both the real parts of $\mu_\mm{nat} (t)$ (dashed
orange trace) and $\mu(t)$ (blue solid trace).

We stress that this behavior is generic for any finite dimension and dressing choice, and is neither specific to the two-mode
model we consider nor the SATD dressing.

\subsubsection{Non-Hermitian SATD Control Hamiltonian}
\label{sec:SATDCtrlW}

In the dressed frame defined by the SATD transformation [see Eq.~\eqref{eq:S_dr_satd}], the non-Hermitian SATD control Hamiltonian
is given by
\begin{equation} 
    W_\mm{dr,SATD} (t) = i \hS_\mm{SATD}^{-1} (t) \rd_t\hS_\mm{SATD} (t) = \frac{1}{2} \dot{\mu} (t) \hat{\sigma}_x,
    \label{eq:W_SATD}
\end{equation}
and is only defined when $\hS_\mm{SATD} (0)=\hS_\mm{SATD} (t_0)=\mathbbm{1}$ [see Eq.~\eqref{eq:STA_condition}] up to a global
phase. Transforming Eq.~\eqref{eq:W_SATD} to the original lab frame yields
\begin{equation} 
    \begin{aligned}
        \hW_\mm{sym,SATD} (t) &= \hS_\mm{ad}\hS_\mm{SATD}  \hW_\mm{dr,SATD}   \hS_\mm{SATD}^{-1}  \hS_\mm{ad}^{-1} \\
        &= g_x (t) \hat{\sigma}_x + g_z (t) \hat{\sigma}_z,
    \end{aligned}
    \label{eq:W_SATD_lab}
\end{equation}
where we omitted the explicit time dependence of the change of frame operators in the first line for simplicity. The control
fields are given by
\begin{equation} 
    \begin{aligned}
        g_x (t) &= -\Omega (t) + \frac{\dot{\theta}(t)}{2}\cot{\mu (t)}\sin{\theta (t)}+\frac{\dot{\mu}(t)}{2}\cos{\theta (t)}\\
        g_z(t) &= \frac{i\Gamma}{2}+\Delta (t)+ \frac{\dot{\theta}(t)}{2}\cot{\mu (t)}\cos{\theta (t)} -
        \frac{\dot{\mu}(t)}{2}\sin{\theta (t)}.
    \end{aligned}
    \label{eq:gx_gz}
\end{equation}

In Sec.~\ref{sec:TD_dynamics}, we stressed that for the two-mode case there was no difference in using $\hS_\mm{ad}(t)$ [see
Eq.~\eqref{eq:S_ad_holo}] or a non-holomorphic version of it to obtain $\hW_\mm{ad,TD} (t)$ [see Eq.~\eqref{eq:W_TD_ad}]. However,
Eq.~\eqref{eq:W_SATD_lab} shows that using the non-holomorphic version of $\hS_\mm{ad}(t)$ leads to an invalid STA. The
$\pi$-phase jumps (see Sec.~\ref{sec:2ModeHolomorphicS}) will result either in $g_x (t)$ and $g_z (t)$ being discontinuous or
having the wrong sign. 

\subsubsection{SATD Dynamics}

As for the Hermitian case, the SATD modified non-Hermitian Hamiltonian $\hH_\mm{sym,SATD} (t) = \hH_\mm{sym}(t) +
\hW_\mm{sym,SATD} (t)$ generates a dynamics that maps an eigenmode $\ket{\psi_i (0)}$ of $\hH_\mm{sym}(t)$ prepared at $t=0$ to
$\ket{\psi_i (t_0)}$ without occupying $\ket{\psi_i (t)}$ at all times. We show this in Figs.~\ref{fig:corr_err_combo}(a) and
\ref{fig:corr_err_combo}(b) where we plot the transition probability $P_{i,i} (t)$ (thin solid purple trace) as a function of time
for $\Gamma_0 t_0 = 5$ for $i \in \{+,-\}$. The transition probability obeys $P_{i,i} (0) = P_{i,i} (t_0) = 1$ while $P_{i,i} (t)
\neq 1$ as expected. The numerics were performed with the same control loop as the one used in Sec.~\ref{sec:TD_dynamics}. 

We also show in Figs.~\ref{fig:corr_err_combo}(c) and (d) the behavior of the state dependent fidelity errors $\mathcal{E}_{i,i}
(t_0)$ as a function of  $\Gamma_0 t_0$ [see Eq.~\eqref{eq:fidelity_error}] for $i \in \{+,-\}$, respectively. Our results show
that for a valid SATD protocol (see Sec.~\ref{sec:SATD_DT}) we can always map $\ket{\psi_i (0)}$ to $\ket{\psi_i (t_0)}$. This
implies that although the dynamics generated by $\hH_\mm{sym,SATD}$ does trace the ideal eigenvalue braids associated to
$\hH_\mm{sym}[\gamma(\varepsilon)]$, it sill allows one to generate the topological permutation operation defined in
Eq.~\eqref{eq:sigma_x}.

Similar to our analysis of the TD correction in Sec.~\ref{sec:TD}, we check if the non-trivial braids traced by $\hH_\mm{sym,SATD}
(t)$ are linked to encircling an EP associated to the modified spectrum. Our results are displayed in Fig.~\ref{fig:STA_SLDL}(b)
where we show as a function of loop time $\Gamma_0 t_0$ and radius $\Delta_0/\Gamma_0$ a map  indicating if an EP of
$\hH_\mm{sym,SATD} (t)$ is encircled or not when a non-trivial braid is traced. We check this using $1\leq\Gamma_0 t_0\leq
50,0.1\leq\Delta_0 t_0\leq 1, \varphi=0,\Omega_0=\Gamma_0/2,$ and $d=1$. Similarly to TD (see Sec.~\ref{sec:TD}), we conclude that
a valid SATD protocol always leads to a non-trivial braid regardless of winding around an EP associated to $\hH_\mm{sym,SATD}
(t)$. 

\subsubsection{Choice of Initial Control Loop}
\label{sec:SATDChoiceInitLoop}

In Sec.~\ref{subsec:phi_eq_pi}, we discussed that there exists choices of base points for the same contour that can lead to the
imaginary part of the spectrum to be anti-symmetric with respect to $t=t_0/2$. Evolving the system along the associated control
loop and fine-tuning the duration of the loop leads to $P_{i,i} (t_0) = 1$ for $i \in \{+,-\}$ [see Fig.~\ref{fig:phieqpi}], but
with $P_{i,i} (t) \neq 1$ for intermediate times. 

\begin{figure}[t] 
    \centering
    \includegraphics[width=\columnwidth]{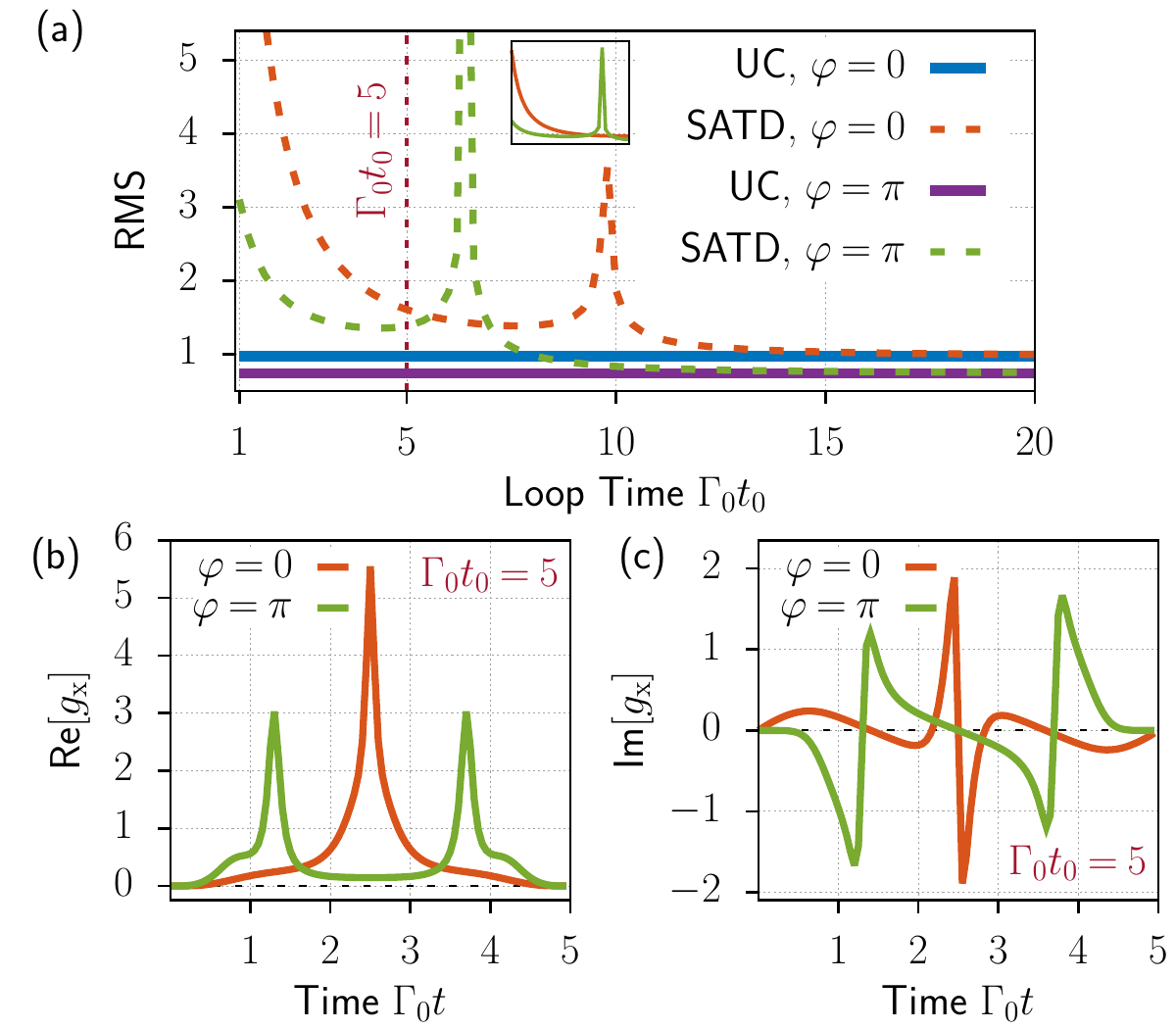}
    \caption{RMS amplitude for SATD correction. (a) Comparison of RMS [see Eq.~\eqref{eq:RMS}] amplitudes between uncorrected case
        (UC, solid lines) and SATD (dashed lines) for $\varphi=0$ (blue and orange traces) and $\varphi=\pi$ (purple and green).
        Example of (b) real and (c) imaginary parts of $g_x (t)$ [see Eq.~\eqref{eq:gx_gz}] for $\Gamma_0 t_0 = 5$. There is no
        appreciable difference for the $g_z(t)$ control field. Choosing $\varphi=\pi$ yields control fields with an overall
        reduced amplitude.
    }
    \label{fig:SATD_phi_eq_pi}
\end{figure}

In Fig.~\ref{fig:SATD_phi_eq_pi}(a) we compare the RMS amplitude [see Eq.~\eqref{eq:RMS}] of the fields needed to implement the
initial (uncorrected, UC) circular control loop (solid traces) [see Eq.~\eqref{eq:Num_delta_Num_omega}] and the associated SATD
protocol (dashed traces) as a function of loop time $\Gamma_0 t_0$ for both $\varphi=0$ (blue and orange traces) and $\varphi=\pi$
(purple and green traces), with the latter choice yielding a spectrum with an anti-symmetric imaginary part with respect to
$t_0/2$. In our numerics, we further use $\Delta_0=\Omega_0=\Gamma_0/2, \varphi=0,$ and $d=1$.

Deriving the SATD correction from a control loop with basepoint $\varphi=\pi$ yields on average a modified protocol that requires
less resources to be implemented. In particular, choosing $\varphi=\pi$ and tuning the loop time $\Gamma_0 t_0$ to yield
quasi-unitary dynamics [see Figs.~\ref{fig:phieqpi}(b) and (c)] leads to SATD control fields that have an amplitude which is more
than 50\% smaller than those obtain with $\varphi=0$. We illustrate this behavior in Figs.~\ref{fig:SATD_phi_eq_pi}(b) and (c)
where we plot the real and imaginary parts of $g_x (t)$ [see Eq.~\eqref{eq:gx_gz}] for $\Gamma_0 t_0 = 14.5$, respectively. The
orange (green) traces in Figs.~\ref{fig:SATD_phi_eq_pi}(b) and (c) correspond to choosing $\varphi =0$ ($\varphi=\pi$).

In contrast to TD (see Sec.~\ref{sec:TDChoiceInitLoop}), the resources needed to implement SATD can largely be reduced by choosing
an initial control loop that yields a quasi-unitary dynamics that almost perfectly maps $\ket{\psi_i (0)}$ to $\ket{\psi_i
(t_0)}$. This simply follows from the SATD philosophy, which does not enforce the ideal adiabatic dynamics at all times, but only
at the initial and final times. Thus, for those special control loops, SATD needs to correct the dynamics ``less'', which
translates into correction fields with smaller amplitudes.

\subsection{Robustness Analysis}
\label{sec:Robustness}

For Hermitian systems, STAs inherit some of the robustness against parameter uncertainties from the adiabatic protocol they
originate from~\cite{Odelin2019,ribeiro2021,ibanez2011,Bason2012,Chen2011}. Since parameter uncertainties are a form of
quasi-static noise, they induce small-amplitude, noise-mediated transitions that are suppressed due to a large energy gap between
instantaneous eigenmodes. To a certain degree, this behavior is an extension of the adiabatic theorem for small-amplitude,
noise-mediated transitions between eigenmodes.

For non-Hermitian systems, however, since the adiabatic theorem does not hold (see Sec.~\ref{subsec:prelim_dynamics}), one might
conclude that STAs offer no intrinsic form of robustness against parameter uncertainties. While this is true in the long loop-time
limit ($\Gamma_0 t_0 \gg 1$), where any small amplitude transitions from the most- to the least-lossy mode are exponentially
amplified (see Sec.~\ref{subsec:dynamics}), it is not the case in the short loop-time limit ($\Gamma_0 t_0 \lesssim 1$) since
noise-mediated transitions are only marginally amplified.

To understand how parameter uncertainties give rise to noise-mediated transitions, it is enough to consider the scenario where
there is an uncertainty in the amplitude of the control fields of $\hH_\mm{sym}$ [see Eq.~\ref{eq:H_sym_static}] generating the
control loop $\gamma_\mm{circ}\{\pv[\varepsilon(t)]\}$. Specifically, we take the parameter $\Delta_0$ [see
Eq.~\eqref{eq:Num_delta_Num_omega}] to incur an uncertainty of $\delta \Delta_0$. Here, without loss of generality, we assume that
$\delta \Delta_0$ is characterized by a normal Gaussian distribution
\begin{equation} 
    p(\delta\Delta_0) = \frac{1}{\beta \sqrt{2\pi}} \exp\left[-\frac{1}{2}\left(\frac{\delta\Delta_0}{\beta}\right)^2\right],
    \label{eq:gaussian}
\end{equation}
with mean $\langle \delta\Delta_0 \rangle_{\delta \Delta_0} = 0$ and variance $\langle \delta\Delta_0^2 \rangle_{\delta \Delta_0}
= \beta^2$, where $\langle \cdot \rangle_{\delta \Delta_0}$ denotes averaging over the stochastic variable $\delta \Delta_0$.

Within this framework, the non-Hermitian Hamiltonian describing the evolution of the system is  $\hH_\mm{sym,imp} (t)=
\hH_\mm{sym} (t) + \delta \Delta_0 \hat{\sigma}_z$, which in the adiabatic frame of $\hH_\mm{sym} (t)$ takes the form
\begin{equation}
    \begin{aligned}
        \hH_\mm{ad,imp}(t) &= \hS^{-1}_\mm{ad} (t) \hH_\mm{sym, imp} (t) \hS_\mm{ad}(t) - i \hS^{-1}_\mm{ad} (t) \rd_t\hS_\mm{ad} (t) \\
        &= \hH_\mm{ad}(t) + \delta\Delta_0 \left( \cos \left[2 \theta(t)\right] \hat{\sigma}_z-  \sin \left[2 \theta(t)\right] \hat{\sigma}_x
        \right),
    \end{aligned}
    \label{eq:H_ad_imp}
\end{equation}
with $\hH_\mm{ad} (t)$ given in Eq.~\eqref{eq:H_ad_dyn}. Equation~\eqref{eq:H_ad_imp} shows that any uncertainty on the parameter
$\Delta_0$ will result in non-adiabatic-like transitions which cannot be corrected with STAs since the value of $\delta \Delta_0$
is unknown.

\begin{figure}[t] 
    \includegraphics[width=\columnwidth]{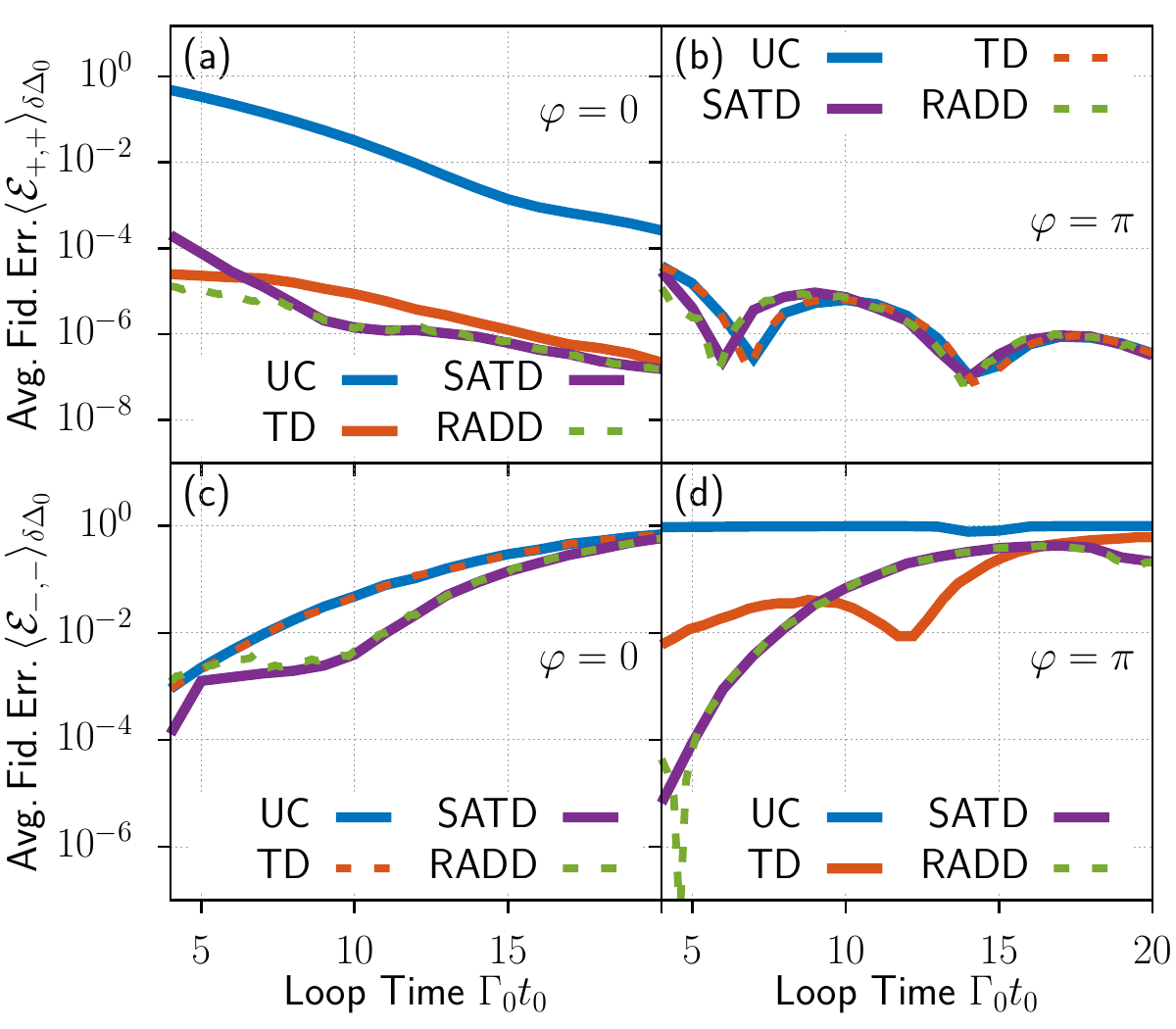}
    \caption{Robustness of STAs against parameter uncertainties. (a) Average state fidelity error
    $\langle\mathcal{E}_{+,+}(t_0)\rangle$ [see Eqs.~\eqref{eq:fidelity_error} and
    Eq.~\eqref{eq:imp_P}] as a function of $\Gamma_0 t_0$ for the circular control loop $\gamma_\mm{circ}$ [see
    Eq.~\eqref{eq:epsilon_time}] using $\varphi=0$. (b) Same as (a) but with $\varphi=\pi$ [see Fig.~\ref{fig:phieqpi} and
    Sec.~\ref{subsec:phi_eq_pi}]. (c) Same as (a) for $\langle\mathcal{E}_{-,-}(t_0)\rangle$. (d) Same as (c) for $\varphi=\pi$.}
    \label{fig:IMP}
\end{figure}

To illustrate the robustness of STAs in the short-time limit, we plot in Fig.~\ref{fig:IMP} the noise-averaged average fidelity
error 
\begin{equation} 
    \begin{aligned}
        &\langle\mathcal{E}_{i,j}(t)\rangle_{\delta \Delta_0} = 1 -\langle P_{i,j}(t) \rangle_{\delta \Delta_0}\\
        &= 1- \int_{-\infty}^{\infty} \ud\delta \Delta_0 p(\delta \Delta_0)\frac{|\bra{\psi_j} \hat\Phi_\mm{ad,imp}(t)\ket{\psi_i}|^2}{\sum_{j}
        |\bra{\psi_j} \hat\Phi_\mm{ad,imp}(t)\ket{\psi_i}|^2},
    \end{aligned}
    \label{eq:imp_P}
\end{equation}
as a function of the loop time $\Gamma_0 t_0$ for the uncorrected circular control loop [see Eq.~\eqref{eq:Num_delta_Num_omega}]
(UC, blue trace) and the associated TD (orange trace) and SATD (purple trace) protocols for both $\varphi=0$ (see
Figs.~\ref{fig:IMP}(a) for $i=j=+$ and (c) for $i=j=-$) and $\varphi = \pi$ (see Figs.~\ref{fig:IMP}(b) for $i=j=+$ and (d) for
$i=j=-$). The numerics were performed by choosing $4\leq\Gamma_0 t_0\leq 50,\Delta_0=\Omega_0=\Gamma_0/2, \varphi=0,$ and $d=1$. 

Our results show the anticipated behavior:~Non-Hermitian STAs provide a certain degree of robustness against parameter
uncertainties in the short loop-time limit and allow one to generate the desired topological operation with high-fidelity for loop
times fulfilling the condition $\Gamma_0 t_0 \lesssim 7$.

In principle, the non-Hermitian versions of TD and SATD should allow one to realize eigenvalue braiding in an experimental setup,
even in the presence of uncertainties, given that one performs a ``fast'' control loop. However, this requires one to realize
control fields with relatively large amplitudes since the corrections fields scale like $1/t_0$ [see Eq.~\eqref{eq:W_TD_ad} for
TD, and Eq.~\eqref{eq:W_SATD} for SATD]. 
 
In the next section, we show how an appropriate choice of dressing leads to control fields with a moderate amplitude even in the
short loop-time limit. This type of STA can be much easier to implement for a given experimental platform.

\subsubsection{Reduced Amplitude Dressed Driving} 
\label{subsubsec:RADD}

We design a dressing that yields control fields with an RMS amplitude [see Eq.~\eqref{eq:RMS}] smaller than TD and SATD in the
short loop-time limit. To this end, we consider a modification of the SATD dressing angle given by 
\begin{equation} 
    \mu_\mm{mod}(t) = - \arctan\left[ \frac{\frac{1}{2}\dot{\theta}(t)}{\lambda(t) [1+\mm{F}(t)]} \right],
    \label{eq:Mod_mu_dr}
\end{equation}
where 
\begin{equation} 
    \mm{F}(t) = \mm{A} \exp{\left[-\left(\frac{t-t_0/2}{\nu}\right)^{2n}\right]},
    \label{eq:mod_gaussian}
\end{equation}
is a hyper-Gaussian mask function with free parameters $A$, $\nu$, and $n$. The purpose of $F(t)$ is to reduce the overall
amplitude of the control fields $g_x (t)$ and $g_z (t)$ [see Eq.~\eqref{eq:gx_gz}] by substituting $\mu (t) \to \mu_\mm{mod} (t)$
since the fields are proportional to the amplitude of $\mu (t)$ and its first derivative. This defines a new STA which we coin
Reduced Amplitude Dressed Driving (RADD).

To minimize the amplitude of $g_x (t)$ and $g_z (t)$ we look for the set of values $\{A, \nu, n\}$ that minimize the RMS amplitude
[see Eq.~\eqref{eq:RMS}] of the control fields for a given $t_0$. The minimization procedure is done by considering finite ranges
for $A$, $n$, and $\nu$. We chose $A \in [10^{-2},10]$, $n \in [1,7]$, and $\nu \in [t_0/25,t_0/3]$.

\begin{figure}[t] 
    \centering
    \includegraphics[width=\columnwidth]{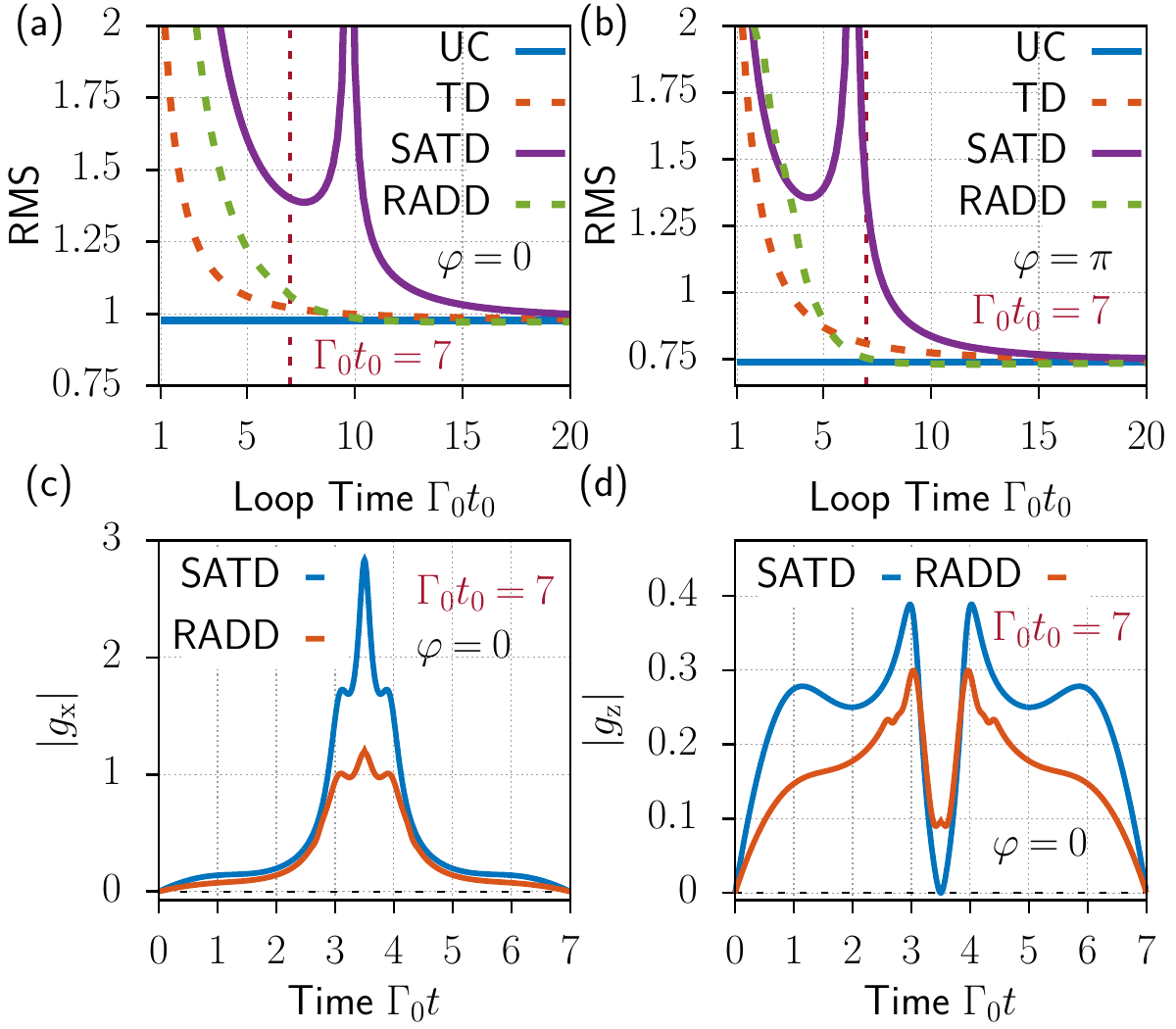}
    \caption{Comparison of resources needed to implement different STAs. (a) RMS amplitude as a function of $\Gamma_0 t_0$ for the circular
        control loop $\gamma_\mm{circ}$ [see Eq.~\eqref{eq:epsilon_time}] with basepoint at $\varphi=0$. (b) Similar to (a), but with
    $\varphi=\pi$. (c) Absolute value of $g_x[i]$ for $i \in \{\mu, \mu_\mm{mod}\}$ [see Eqs.~\eqref{eq:gx_gz}, \eqref{eq:mu_dr_holo},
    and \eqref{eq:Mod_mu_dr}] showing the overall improvement in field magnitude for $\Gamma_0 t_0 =10$ and $\varphi=0$. (d) Same as (c)
    but for $g_z$. In (c), we plotted $W_\mm{TD}$ for comparison [see Eq.~\eqref{eq:W_TD_ad}].} 
    \label{fig:RATD}
\end{figure}

In Fig.~\ref{fig:RATD}(a), we plot the RMS amplitude as a function of loop time $\Gamma_0 t_0$ for the uncorrected fields of
$\hH_\mm{sym} (t)$ (UC, blue trace) [see Eq.~\eqref{eq:H_sym_static}] and the associated TD (dashed orange trace), SATD (solid
purple trace), and RADD (dashed green trace) STAs. The numerics were performed with $1\leq\Gamma_0 t_0\leq
20,\Delta_0=\Omega_0=\Gamma_0/2,$ and $d=1$. 

As our results show, there is a range of control loop times $\Gamma_0 t_0 \gtrsim 10$ for which the control fields associated to
the RADD dressing have an RMS amplitude smaller than those of TD and SATD. For $\Gamma_0 t_0=7$, the upper-bound limit of loop
times we found that yields robust STAs in spite of parameter imprecisions [see Fig.~\ref{fig:IMP}], the RADD RMS value is similar
to TD and UC, and 25\% smaller than SATD. Similarly to SATD, RADD allows one to trace an eigenvalue braid that is not necessarily
equivalent to the eigenvalue braid defined by $\hH_\mm{sym}$, but still fulfills Eqs.~\eqref{eq:braiding}, thus, generating the
desired topological operation [see Eq.~\eqref{eq:sigma_x}]. 

We can further reduce the RMS amplitude of the control fields by choosing the base point of the control loop such that
$\mm{Im}[\lambda_\pm (t)]$ is anti-symmetric with respect to $t=t_0/2$, i.e., setting $\varphi = \pi$ and $\Omega_0=\Gamma_0/2$
[see Fig.~\ref{fig:phieqpi} and Sec.~\ref{subsec:phi_eq_pi}]. We compare in Fig.~\ref{fig:RATD}(b) the RMS amplitudes of the
fields when $\varphi=\pi$. At $\Gamma_0 t_0=7$, the RADD RMS (dashed green line) is similar to that of the UC fields (blue trace)
and TD (dashed orange trace), and 45\% smaller than SATD (solid purple trace). Moreover, we find that the RADD RMS value for
$\varphi = \pi$ is $25\%$ smaller compared to the RADD RMS value for $\varphi=0$.

As a consequence of reducing the RMS amplitude, RADD also reduces the overall amplitude of the control fields. To highlight the
improvement in the field amplitudes, we compare in Figs.~\ref{fig:RATD}(c) and (d) the absolute value of $g_x (t)$ and $g_z (t)$
between RADD (orange trace) and SATD (blue trace) for $\Gamma_0 t_0=7$ and $\varphi=0$, respectively.  Our results show that RADD
reduces the overall field amplitudes, while yielding low average fidelity errors. For the chosen loop time, $\Gamma_0 t_0=7$, we
have an average fidelity error of $\langle \mathcal{E}_{-,-}\rangle \lesssim 10^{-3}$ [see Fig.~\ref{fig:IMP}(c)], associated to
generating the topological defined in Eq.~\eqref{eq:sigma_x}.

We stated that $\Gamma_0 t_0=7$ is the upper-bound limit for loop times to be robust against parameter imprecisions, but as our
results indicate it is also the lower-bound limit that yields control fields with ``reasonable'' amplitudes. Since the control
field amplitude of TD scales like $1/t_0$ [see Eq.~\eqref{eq:W_TD_ad}] and that of SATD and RADD like $1/t_0^2$ [see
Eq.~\eqref{eq:gx_gz}], one cannot arbitrarily shorten the control loop time. This would result in impractical control fields.  We
surmise this behavior to be linked to the existence of a speed limit akin the quantum speed limit~\cite{Mandelstam1991}.

\subsection{Experimental Realization}

We now provide an example of an experimental platform, where our control schemes can be realized.

We consider an optomechanical setup with two laser drives, such that each mechanical mode is coupled to two optical
modes~\cite{xu2016}. Tracing out the degrees of freedom associated to the optical modes leads to the effective non-Hermitian
Hamiltonian given by 
\begin{equation} 
    \hH= \sum_{j=1}^2 \left(\omega_{\mm{mech},j}-i\frac{\gamma_{\mm{mech},j}}{2}-ig_j^2\eta\right) \ketbra{j}{j} - i \eta g_{1}g_{2}\hat{\sigma}_x.
    \label{eq:real_H}
\end{equation}
where $\omega_{\mm{mech},j}$,  $\gamma_{\mm{mech},j}$ and  $g_{j}$ are the frequency, damping rate, and optomechanical coupling
constant of the $j$th mechanical mode, respectively. The complex mechanical susceptibility $\eta$ is given by
\begin{equation} 
    \begin{aligned}
            \eta = & \frac{P_\mm{L}}{\hbar \Omega_{\mm{L},j}} \frac{\kappa_{\mm{in},j}}{(\kappa/2)^2+\delta_0^2} \times \\
            & \left[\frac{1}{\kappa/2-i(\omega_0+\delta_0)}-\frac{1}{\kappa/2+i(-2\omega_0+\delta_0)}\right],
    \end{aligned}
    \label{real_sigma}
\end{equation}
where $\delta_0$ is the mean detuning between the laser and the cavity, $\kappa$ is the linewidth of the cavity, $\kappa_\mm{in}$
is the input coupling rate of the cavity, $\omega_0=(\omega_1+\omega_2)/2$, $P_\mm{L}$ is the power of the laser and
$\Omega_\mm{L}$ its frequency. Equation~\eqref{eq:real_H} can be made time-dependent by making both the power and frequency of
the laser time dependent. Within this framework, the effective NH Hamiltonian describing the evolution of the optomechanical setup
resembles the two-mode NH Hamiltonian defined in Eq.~\eqref{eq:H_sym_static}.

By mapping Eq.~\eqref{eq:real_H} to $\hH_\mm{sym} (t) + \hH_\mm{RADD} (t)$ [see Eqs.~\eqref{eq:H_sym_static} and
\eqref{eq:Mod_mu_dr}], we can find the modified time-dependent optomechanical NH that allows one to generate the desired braiding
operations. This procedure reduces to solving two equations for two unknowns: $P_\mm{L} (t)$ and $\delta_0 (t)$. 

\section{Conclusion}

In this work we have demonstrated how to construct shortcuts-to-adiabaticity for non-Hermitian systems in the presence of spectral
singularities. Failing to account for the latter and naively following the recipe developed for Hermitian Hamiltonians will lead
to control schemes that do not generate the desired dynamics. 

We found that valid NH STAs, in stark contrast to the Hermitian case, do not lead to a family of control schemes characterized by
the protocol duration. Besides the fundamental aspect, this has practical implications since a given NH STA may be constraint to
long protocol durations. As we have discussed, NH STAs are unstable in this regime in the presence of parameter
uncertainties.

The lack of inherent robustness limits the use of celebrated protocols like TD and SATD in realistic settings, but as we have
shown the versatility of our approach allows one to find STAs that are robust against parameter uncertainties and constrained
in terms of experimental resources. Our results pave the way towards a new class of control protocols, for both classical and
quantum systems, that leverage the interplay between coherent and incoherent dynamics. 

\appendix

\section{Criteria for Eigenvalue Braiding} 
\label{sec:app_eigeval_braid}

For a two-mode system described by a symmetric NH Hamiltnian, i.e., $\mm{Tr}[\hH (t)] = 0$, the following conditions hold true:
\begin{equation}
    \begin{aligned}
        \text{(I)}:\,  &\lambda_\pm (0) = \lambda_\mp (t_0) = \lambda_\pm (2t_0),\\
        \text{(II)}:\, &\int_0^{t_0} \di{t} \lambda_\pm (t)  \neq 0 \,\text{ and,}\\ 
                       &\int_0^{2t_0} \di{t} \lambda_\pm (t) = 0.
    \end{aligned}
    \label{eq:ep_encircle_braiding_conditons}
\end{equation}
where $\lambda_\pm (t)$ are the holomorphic eigenvalues associated to the NH Hamiltonian $\hH (t)$. 

Condition (I) expresses the fact that encircling an $\mm{EP}_2$ twice with a control loop that results from the concatenation of
two loops with the same orientation brings the spectrum back to its original position on the Riemann surface associated to the
spectrum of $\hH (t)$. 

Condition (II) follows from condition (I) and expresses the fact that encircling an $\mm{EP}_2$ twice with a control loop that
corresponds to the concatenation of two identical loops leads to $\lambda_\pm (t)$, $t\in[0,2t_0]$, to be an anti-symmetric
function with respect to $t=t_0$. 
\vspace{1.5cm}

\end{document}